\documentclass[%
 reprint,
 amsmath,amssymb,
 aps,
]{revtex4-2}

\usepackage{graphicx,xcolor,comment}
\usepackage{dcolumn}
\usepackage{bm}
\usepackage{array}
\usepackage[caption=false]{subfig} % <-- replace subcaption with this
\usepackage[utf8]{inputenc}
\usepackage{textcomp}   % for the minus sign

\newcommand{\Msun}{\mathrm{M}_\odot}

\newcommand{\mdot}{\dot{m}}

\begin{document}

\preprint{}

\title{G objects as Primordial Black Hole-Neutron Star Remnants:\\ Population Modeling and Multi-Wavelength Observables}% Force line breaks with \\

\author{David Morales-Zapien}%
 \email{David.Morales-Zapien02@student.csulb.edu}
\affiliation{%
 Department of Physics \& Astronomy, California State University Long Beach, Long Beach, CA 90840, U.S.A.
}%

\author{Stefano Profumo}%
 \email{profumo@ucsc.edu}
\affiliation{Department of Physics, 
University of California Santa Cruz, Santa Cruz, CA 95064, USA}
\affiliation{Santa Cruz Institute for Particle Physics,
Santa Cruz, CA 95064, USA}


\begin{abstract}
The nature of the so-called G objects orbiting the Galactic Center remains unresolved. These sources exhibit compact Br$\gamma$ emission, extreme infrared colors, and remarkable dynamical stability through close passages to the central supermassive black hole, challenging conventional interpretations as stars or unbound gas clouds. We investigate the hypothesis that G objects are the remnants of neutron stars that have been converted into low-mass black holes through the capture of primordial black holes, a viable dark-matter candidate. We construct a population-level framework linking the abundance and spatial distribution of these remnants to the neutron-star population, the inner dark-matter density profile, and the primordial black-hole mass and abundance. Within this framework, the observed G-object population arises naturally from the same PBH-induced neutron-star conversion process, while remaining consistent with the observed deficit of ordinary radio pulsars in the Galactic Center, which is not significantly affected by this mechanism. We further identify a suite of observational signatures—across infrared, radio, X-ray, and microlensing channels—that render this scenario empirically testable and distinguishable from stellar-envelope models. Our results show that G objects can act as sensitive probes of compact-object capture physics and of dark matter on sub-galactic scales.\end{abstract}

\maketitle

%\tableofcontents

\section{\label{sec:Intro}Introduction}

The nature of the so-called G objects in the Galactic Center, initially identified as a pair of sources, most prominently one identified as ``G2'', and subsequently expanded by the discovery of six additional objects, remains one of the most intriguing open problems in Galactic astrophysics \cite{Ciurlo:2020Nature, Gillessen2012, Gillessen:2013G2,Phifer_2013, Cl_net_2005, Pfuhl_2015}. Observationally, this class is defined by the presence of distinct and spatially compact Br$\gamma$ emission, weak near-infrared $K$-band continuum relative to $L'$ emission ($K'-L' \gtrsim 4.5$), and large proper-motion and radial-velocity variations over time, indicating bound Keplerian orbits about the central supermassive black hole, Sgr~A$^*$ \cite{Ciurlo:2020Nature}. By compact, we refer to sources that are {\em unresolved} ($\lesssim 0.03''$) or only {\em marginally} resolved ($\sim 0.05''$) at the distance of the Galactic Center. Despite repeated periapse passages at distances of only a few thousand Schwarzschild radii, they remain dynamically intact, firmly excluding interpretations as unbound gas clouds.

The survival of G objects through periapse~\cite{Witzel:2014G2,Valencia_S__2015,Plewa:2017G2}, together with their compactness and orbital stability, implies the presence of a  gravitationally bound core, accompanied by a gaseous and dusty envelope. This fact has motivated a variety of stellar interpretations, including dust enshrouded stars ~\cite{Peissker2020, Peissker2025}, young stellar objects with circumstellar disks~\cite{Murray_Clay_2012, Owen_2022}, planetary embryos~\cite{mapelli2015}, stellar merger remnants~\cite{Witzel:2014G2, Prodan_2015,Stephan2016MergingBinariesGC, Stephan2019}, stellar collisions~\cite{rose2023} and partially stripped stars \cite{Phifer_2013, gibson2024}. While such models can reproduce individual observational features, open questions remain regarding the availability of progenitors ~\cite{Owen_2022} and formation channel ~\cite{Ciurlo:2020Nature}, the survival of compact dusty envelopes in the Galactic Center environment ~\cite{zajacek2024galacticcentergobjects}, and the absence of an unambiguous stellar signature from the known members of the class ~\cite{Gillessen2012,Plewa:2017G2, Pfuhl_2015}.

The interpretation of the G objects as systems containing compact, gravitationally bound cores does not uniquely determine the physical nature of those cores, motivating consideration of scenarios beyond conventional stellar remnants. In this context, it is natural to consider the possible role of dark matter, whose fundamental nature and spatial distribution in the Galactic Center remain poorly constrained. Dark matter has been widely explored in the form of weakly interacting  particles, such as WIMPs~\cite{Bertone2005, Jungman1996, LZ2025PRL}, axions \cite{marsh2016} or sterile neutrinos \cite{Dodelson1994} which populate a dark sector and interact only feebly with baryonic matter. However, the longstanding core--cusp problem~\cite{Bullock_2017, de_Blok_2009, Genina_2017} leaves open the possibility that the dark-matter density in the central regions of the Galaxy may be sufficiently enhanced for alternative dark-matter realizations to become astrophysically relevant. In particular, dark matter may be macroscopic in nature, consisting of rare but massive objects with large interaction cross sections that evade detection due to their low number density. Primordial black holes (PBHs) provide a well-motivated realization of such a macroscopic dark matter scenario and are hypothesized to have formed in the early universe through the collapse of large primordial density perturbations \cite{hawking1971, Carr_2021, zeldovich1967, chapline1975l}. Such overdensities may arise from inflationary dynamics \cite{Cotner_2017, Cotner_2018, Inomata_2017}, cosmological phase transitions \cite{khlopov1980primordial}, or the collapse of topological defects \cite{crawford1982spontaneous}.

In sufficiently dense dark-matter environments, PBHs can participate in compact-object interactions that are otherwise negligible. In particular, a PBH traversing a neutron star can lose kinetic energy through dynamical friction and accretion, become gravitationally bound, settle to the stellar core, and subsequently accrete neutron-star material \cite{Flores:2023PRD, Takhistov_2021, Fuller_2017, Caiozzo_2024, G_nolini_2020}. This process leads to the destruction of the neutron star and the formation of a low-mass ($\sim$1--2~$M_\odot$) black hole on timescales short compared to Galactic ages \cite{Flores:2023PRD, Takhistov_2021, Fuller_2017, Caiozzo_2024, G_nolini_2020}.

In this work, we investigate in detail the hypothesis, originally proposed in Ref.~\citet{Flores:2023PRD}, that  { G objects are the remnants of neutron stars converted into black holes through the capture of PBHs}. In this scenario, the compact core of a G object is a stellar-mass black hole rather than a star, while the observed infrared continuum and recombination-line emission arise from a compact envelope or a halo of gas and dust gravitationally bound to the compact remnant. Such objects are intrinsically resistant to tidal disruption by Sgr~A*, naturally explaining their survival at periapse, while their gaseous envelopes provide the characteristics required to reproduce the observed infrared phenomenology.

Our approach is twofold. First, we develop a population-level framework that connects the abundance and radial distribution of G objects to the underlying neutron-star population, the Galactic dark-matter density profile, and the fraction and mass of PBHs. In this picture, the observed G objects and the apparent deficit of ordinary radio pulsars in the Galactic Center were initially considered complementary outcomes of PBH-induced neutron-star conversion; however, our results show that this mechanism is insufficient to significantly deplete the pulsar population. Second, we identify a set of multi-wavelength observational diagnostics—including infrared, radio, X-ray and microlensing signatures—that can distinguish compact, gas-enshrouded black-hole remnants from conventional stellar-envelope models on both individual and population scales.

The aim of this paper is not to argue that G objects  must arise from PBH interactions, but, rather, to demonstrate that this scenario is  internally consistent, observationally constrained, and  empirically falsifiable. By linking Galactic Center phenomenology to compact-object capture physics and dark-matter structure, we show that G objects can serve as novel probes of both neutron-star populations and the small-scale properties of dark matter.

\section{G objects: Observational Status}
\label{sec:observational_status}

Recent observational campaigns of the objects known as ``G objects'' (notably G1--G6 and especially the prototypical G2) have provided strong empirical constraints that shape---and in many cases narrow---the viable theoretical interpretations of their nature. Below we summarize the key findings and discuss how they inform the distinction between classical dust-enshrouded-star or merger models and alternative compact-object (e.g., PBH + gas halo) hypotheses.

\subsection{Population, infrared emission, and orbital demographics}

High-resolution, multi-epoch adaptive-optics imaging and integral-field spectroscopy (Keck, VLT/GRAVITY) have established that G objects constitute a small but genuine population rather than a single anomalous event \cite{Gillessen2012,Gillessen2013,Witzel2014,Witzel2017}. At least six confirmed members (G1–G6) exhibit compact infrared continuum emission combined with spatially resolved Br$\gamma$ line structure.

Their orbital parameters span a wide range of eccentricities and periapse distances, demonstrating that they are not fragments of a single disrupted cloud. Instead, each object follows an independent Keplerian orbit, implying the presence of a gravitationally bound central mass. The diversity of orbital orientations and energies disfavors common-origin scenarios tied to a single infall event.

Infrared spectral energy distributions indicate dust temperatures of order several hundred Kelvin, consistent with compact dusty envelopes heated either by an embedded stellar source or by external irradiation from the intense ultraviolet field of the central cluster \cite{Witzel2017,Peissker2024DustySCluster}. The recombination-line luminosities cluster around $L_{\rm Br\gamma} \sim 10^{30}$–$10^{31}\ \mathrm{erg\ s^{-1}}$, suggesting envelope gas masses significantly below those of classical star-forming disks but larger than expected for tenuous stellar winds alone.

The existence of dusty S-cluster objects with similar colors but varying gas fractions further supports a continuum interpretation in which G objects represent an extreme subset of dust-enshrouded compact systems rather than a fundamentally distinct class \cite{Peissker2024DustySCluster}.

\subsection{Pericenter passage of G2: morphology, drag, and accretion non-events}

The pericenter passage of G2 in 2014 ($\sim$2000--3000 Schwarzschild radii from Sgr~A$^*$) offered a critical test of competing models. Integral-field spectroscopy revealed significant tidal stretching of the Br$\gamma$ line emission, consistent with gas being stripped or sheared from an extended envelope. However, the object remained compact in $L'$-band continuum and exhibited no dramatic or persistent brightening of Sgr~A$^*$ across radio, millimeter, infrared, X-ray, or TeV wavelengths. This strongly argues against a scenario in which a large amount of unbound gas was dumped onto the supermassive black hole (SMBH) \cite{Gillessen:2013G2,Witzel:2014G2,Plewa:2017G2}.

Subsequent analyses of the orbit and emission indicate only weak, if any, hydrodynamic drag and place stringent upper limits on the density of the ambient accretion flow at radii of order $10^{3}$--$10^{4}\,R_{\rm S}$. These constraints disfavor large, low-density gas clouds and favor objects with small effective cross-sections, such as bound stars with dusty shells or compact halos around massive cores \cite{Plewa:2017G2,Hosseini:2020S2density,Gillessen:2018Drag}.

\subsection{Incompleteness of the Galactic-Center G objects Census}
\label{sec:Gobject_incompleteness}

The currently identified population of Galactic-center G objects
(G1--G6) is almost certainly incomplete. While near-infrared observations
penetrate the heavy foreground extinction toward Sgr~A$^*$, the detection
and classification of G objects are subject to a combination of strong
selection effects. Here, we quantify these effects and estimate
the fraction of the underlying population that is plausibly detectable
with existing data, and attempt to infer the size of the entire population given the associated observational incompleteness.

\subsection{Extinction and line-of-sight geometry}

At near-infrared wavelengths ($K$ and $L'$ bands), the extinction toward
the Galactic Center is dominated by foreground dust, with typical values
$A_K \simeq 2.5$--3.0 and $A_{L'} \simeq 1.0$--1.5
\cite{Fritz:2011Extinction,Schodel:2010Extinction}.
Because most of this extinction arises several kiloparsecs in front of the
nuclear star cluster, sources located slightly behind Sgr~A$^*$ do not
experience dramatically larger attenuation than those on the near side.
Consequently, extinction alone does not prevent the detection of G objects
on the far side of the Galactic Center.

However, far-side sources are projected against a denser stellar background
and bright diffuse emission from ionized gas, which significantly increases
confusion noise. As a result, the practical limitation is not extinction but
source confusion and contrast.

%\subsection{Crowding and confusion}

The stellar surface density within the central arcsecond exceeds
$10^6\,\mathrm{pc^{-2}}$, making source confusion the dominant limitation
for faint or marginally resolved objects
\cite{Schodel:2014NSCReview}.
G objects are identified not only by their continuum emission but also by
their recombination-line signatures, particularly Br$\gamma$.
Against the dense background of stars and ionized gas, only objects with
sufficiently compact morphology and distinct kinematic signatures can be
reliably extracted.

Crowding effects are especially severe for objects whose projected positions
lie close to bright early-type stars or along lines of sight with strong
mini-spiral emission. This introduces a geometric selection bias that disfavors
a substantial fraction of randomly oriented orbits.

\subsubsection{Orbital-phase bias}

A critical and often underappreciated selection effect arises from orbital
phase. The known G objects are preferentially identified near pericenter,
where several factors enhance detectability:

(i) Br$\gamma$ luminosity can increase due to compression and enhanced
photoionization,

(ii) tidal shear produces elongated position--velocity structures that
distinguish G objects from normal stars, and

(iii) large line-of-sight velocities separate their emission from the bulk
of the background gas.

For the highly eccentric orbits inferred for G1 and G2 ($e \gtrsim 0.8$),
the fraction of the orbital period spent within a factor of a few of the
pericenter distance is only
\begin{equation}
  f_{\rm peri} \equiv \frac{\Delta t_{\rm peri}}{P}
  \sim 0.05\text{--}0.15,
\end{equation}
depending on the exact eccentricity and pericenter definition
\cite{Gillessen:2013G2,Plewa:2017G2}.
Thus, even in the absence of other selection effects, only $\mathcal{O}(10\%)$
of the population is expected to be observed during the phase of maximum
detectability.

\subsubsection{Kinematic selection and line confusion}

Identification of G objects relies heavily on the detection of Br$\gamma$
emission with velocities that can be cleanly separated from the background
ionized gas. Objects whose line-of-sight velocities overlap with the bulk
of the mini-spiral or other diffuse components are significantly harder to
identify, even if intrinsically luminous
\cite{Ciurlo:2020Nature}.
This kinematic selection further reduces completeness, particularly for
objects on orbits with low projected radial velocities at the epochs of
observation.

\subsubsection{Luminosity and envelope-mass thresholds}

The observed G objects cluster around Br$\gamma$ luminosities
$L_{\rm Br\gamma} \sim 10^{30}$--$10^{31}\,\mathrm{erg\,s^{-1}}$.
Objects with smaller gas envelopes or more compact halos may still exist dynamically but fall below current detection thresholds in recombination-line emission. This introduces an intrinsic luminosity bias that preferentially selects systems with relatively large gas masses or extended  envelopes
\cite{Gillessen:2013G2,Witzel:2014G2}.

In scenarios where G objects are associated with compact cores (stellar or black-hole) that only intermittently or partially retain gas, this luminosity bias can be substantial.

\subsection{Combined completeness estimate}
\label{combined_incompleteness}
The above effects can be combined into a rough but physically motivated completeness estimate. Conservative representative factors are:
\begin{itemize}
  \item Orbital-phase selection: $f_{\rm peri} \sim 0.05$--0.15;
  \item Projection and crowding losses: $\sim 0.5$;
  \item Kinematic and luminosity selection: $\sim 0.3$--0.5.
\end{itemize}
Multiplying these contributions yields an overall detectable fraction
\begin{equation}
  f_{\rm det} \sim (0.05\text{--}0.15)\times 0.5 \times (0.3\text{--}0.5)
  \sim \text{few}\times10^{-2}.
\end{equation}

We therefore conclude that it is entirely plausible that current surveys are sensitive to only $\sim$1--5\% of the true G objects population within the central $\sim$0.1\,pc. The six currently confirmed G objects could thus represent a parent population larger by one to two orders of magnitude.

Such a level of incompleteness is consistent with the known limitations of crowded-field spectroscopy in the Galactic Center and should be borne in mind when using the observed sample to constrain formation scenarios. Interpreting the observed G objects census therefore requires a population-level framework that connects the underlying parent population to the small fraction of objects that are observationally accessible. 

If G objects arise from neutron stars that have undergone conversion to black holes via PBH capture \cite{Flores:2023PRD}, the relevant parent populations are (i) the Galactic neutron-star and (ii) the dark matter density distribution. In the following section, we construct models for the neutron-star population in the inner Galaxy that will serve as the foundation for our subsequent population and likelihood analyses.

%%%%%%%%%%%%%%%%%%%%%%%%%%%%%%%%%%%%%%%%%%%%%%%%%%%%%%%%%%

\section{PBH capture by Neutron Stars}
\label{sec:PBHcapture}

Before proceeding to a quantitative treatment, we outline the basic physical picture underlying our working hypothesis. We consider the possibility that a subset of G objects are the end products of primordial black hole (PBH) capture by neutron stars in the Galactic Center. In this scenario, a PBH traversing a neutron star loses kinetic energy through dynamical friction, becomes gravitationally bound, and sinks to the stellar center, ultimately consuming the neutron star on a timescale short compared to Galactic ages. The result is a low-mass black hole with mass of order the original neutron-star mass. If residual debris, fallback material, or externally supplied gas remains bound to the compact remnant, it may form a compact dusty envelope capable of producing the observed infrared excess and recombination-line emission. This mechanism differs qualitatively from stellar merger or stripped-star models in that the compact core is no longer a luminous star but a newly formed black hole, implying distinct expectations for mass-to-light ratios, variability, and population statistics. In what follows, we formalize this picture and derive the expected present-day volumetric rate of such PBH-induced neutron-star conversions in the Galactic Center.

\subsection{Neutron-star distribution models}
\label{sec:NSmodels}

A key input to any population-level prediction of neutron-star conversion in
the Galactic Center is the underlying spatial distribution of neutron stars. We adopt the standard and reasonable assumption that the present-day neutron-star distribution follows the observed Galactic pulsar population. Pulsar spatial distributions are empirically calibrated on kiloparsec scales and provide the primary observational handle on the Galactic neutron-star population. We normalize this pulsar-tracing spatial distribution to the global neutron-star abundance inferred from the Galactic core-collapse supernova rate, corresponding to a total population $N_{\rm NS,tot} \sim 10^{8} - 10^9$\cite{Diehl:2006, Keane:2008}.

Analytic models commonly used to describe the Galactic pulsar population—such as
exponential disk and Einasto-like profiles—provide an adequate description of the
large-scale ($\gtrsim\,$kpc) structure of the Milky Way
\citep{Yusifov:2004, FaucherGiguere:2006}. However, these models are
empirically calibrated on kiloparsec scales and are not designed to capture the extreme stellar overdensity observed in the central parsec of the Galaxy. As we demonstrate below, straightforward extrapolations of such models severely under-predict the neutron-star population in the Galactic Center.

To make this limitation explicit, we begin by adopting a minimal two-component Galactic model consisting of a stellar disk and a bulge, assuming that neutron stars trace the pulsar-calibrated stellar components. This disk+bulge framework provides a useful baseline for assessing the neutron-star content of the inner Galaxy.

The stellar disk is modeled in Galacto-centric cylindrical coordinates $(R,z)$ as a radially Gaussian distribution with an exponential vertical profile
\citep{Eckner_2018, Vanhollebeke_2009, Ploeg_2017},
\begin{equation}
\rho_{\rm disk}(R,z) =
\frac{M_{\rm disk}}{4\pi \sigma_r^2 z_0}\,
\exp\!\left(-\frac{R^2}{2\sigma_r^2}\right)
\exp\!\left(-\frac{|z|}{z_0}\right),
\end{equation}
where $\sigma_r = 4500$ pc and $z_0 = 410$ pc denote the radial scale length and vertical scale height, respectively with the mass is listed in Table ~\ref{tab:mw_masses}.

Although the Galactic bulge exhibits a boxy/peanut-shaped morphology
\citep{WeggGerhard2013}, the large natal kicks imparted to neutron stars tend to wash out this structure, yielding a significantly rounder pulsar distribution
\citep{Ploeg_2021}. We therefore adopt a  spherically symmetric bulge profile,
\begin{equation}
\rho_{\rm bulge}(r) =
\rho_{0,\rm bulge}\,
\exp\!\left(-\frac{r^2}{R_m^2}\right)
\left(1+\frac{r}{r_0}\right)^{-1.8},
\end{equation}
where $r=\sqrt{R^2+z^2}$ with characteristic scales
$R_m=1.9\,\mathrm{kpc}$ and $r_0=100\,\mathrm{pc}$.

While the disk+bulge model successfully reproduces the large-scale pulsar and stellar distributions of the Galaxy, both components are intrinsically shallow toward the Galactic Center. The disk density vanishes at small radii due to its radial Gaussian form, while the bulge profile exhibits only a mild central rise. As a result, the combined neutron-star density remains strongly suppressed in the inner Galaxy.

\begin{figure}[t]
  \centering
  \includegraphics[width=0.95\columnwidth]{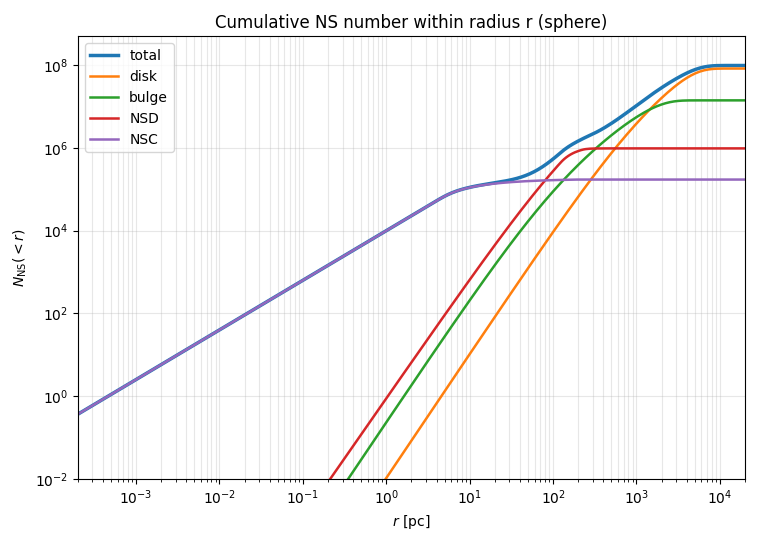}
  \caption{Cumulative neutron-star count predicted by a pulsar-calibrated 4 component Galactic model, normalized to a total
  neutron-star population $N_{\rm NS,tot}\sim10^{8}$.}
  \label{fig:ns_disk_bulge}
\end{figure}

Quantitatively, extrapolating this two-component model into the central parsec predicts only $\mathcal{O}(1$--$10)$ neutron stars within $r<1\,\mathrm{pc}$, depending on the overall NS normalization
with essentially no neutron stars expected inside $0.1\,\mathrm{pc}$
(see Fig.~\ref{fig:ns_disk_bulge}). This is incompatible with the pronounced stellar density enhancement observed in the Galactic Center.% and is insufficient for processes that depend sensitively on the compact-object population at sub-parsec scales such as PBH - NS conversions. 

The failure of a simple disk+bulge description to reproduce the inner Galactic structure thus motivates the inclusion of a distinct nuclear stellar component. Observations of the Milky Way reveal a prominent nuclear stellar overdensity, consisting of both a flattened nuclear stellar disk (NSD) and a dense nuclear
star cluster (NSC), which together dominate the stellar mass and compact-object content within the central hundreds of parsecs while leaving the large-scale Galactic structure largely unchanged \citep{Gallego_Cano_2017, Lauer_2007}. We therefore define a nuclear bulge component as the sum of these two contributions,
\begin{equation}
\rho_{\rm nuc}(R,z) = \rho_{\rm NSD}(R,z) + \rho_{\rm NSC}(r),
\end{equation}
and write the total neutron-star density as
\begin{equation}
\rho_{\rm tot}(R,z) =
\rho_{\rm disk}(R,z) + \rho_{\rm bulge}(r) + \rho_{\rm nuc}(R,z),
\end{equation}
where $r=\sqrt{R^2+z^2}$.

The nuclear stellar disk (NSD) is modeled as a flattened, axisymmetric component distinguished from both the Galactic disk and bulge by its strong radial and vertical stratification. We describe its neutron-star density using the commonly adopted piecewise profile \cite{bartel2025,ploeg2021galactic, Bartels_2018} , which captures the shallow inner slope and steep outer decline associated with the NSD:
\begin{equation}
\rho_{\rm NSD}(r,z)=
\end{equation}
\begin{equation}
\begin{cases}
\rho_{0,\rm NSD}
\left(\dfrac{r}{1~\mathrm{pc}}\right)^{-0.1}
e^{-|z|/45~\mathrm{pc}},
& r < 120~\mathrm{pc}, \\[6pt]
\rho_{1,\rm NSD}
\left(\dfrac{r}{1~\mathrm{pc}}\right)^{-3.5}
e^{-|z|/45~\mathrm{pc}},
& 120~\mathrm{pc} \le r < 220~\mathrm{pc}, \\[6pt]
\rho_{2,\rm NSD}
\left(\dfrac{r}{1~\mathrm{pc}}\right)^{-10}
e^{-|z|/45~\mathrm{pc}},
& r \ge 220~\mathrm{pc}.\label{eq:NSDpiecewise}
\end{cases}
\end{equation}
%
%In the equation above, $r$ is the cylindrical radius and $z$ is the height above the Galactic plane. 
In this form, the NSD exhibits a shallow inner radial dependence ($\propto r^{-0.1}$) transitioning to progressively steeper declines at larger radii, with an exponential vertical scale height of $45~\mathrm{pc}$.

While this component significantly contributes to the stellar and compact-object density on scales of $\mathcal{O}(10$–$100)\,\mathrm{pc}$, its shallow inner slope and finite vertical extent mean that the NSD density rises only weakly toward the Galactic Center. As a consequence, within the central parsec the stellar density of the nuclear star cluster exceeds that of the NSD by orders of magnitude, rendering the NSD contribution negligible for processes localized to the immediate Galactic Center.

\subsection{Nuclear star cluster}
\label{NSC}
The nuclear star cluster is modeled using a Nuker profile, which provides a smooth broken--power-law description of the stellar density in the Galactic nucleus \citep{Gallego_Cano_2017, Lauer_2007},
\begin{equation}
\rho_{\rm NSC}(r) \propto
\left(\frac{r}{r_b}\right)^{-\gamma}
\left[1+\left(\frac{r}{r_b}\right)^{\alpha}\right]^{(\gamma-\beta)/\alpha},
\end{equation}
where $\gamma$ and $\beta$ are the inner and outer logarithmic slopes, respectively, $r_b$ is the break radius, and $\alpha$ controls the sharpness of the transition. Observational fits yield
$r_b \simeq 4.9 \,\mathrm{pc}$ and
$\beta \simeq 3.7 $.

At radii $r \ll r_b$, the nuclear star cluster (NSC) asymptotically approaches a single power-law density profile, $\rho_{\rm NSC}(r) \propto r^{-\gamma}$, analogous to the canonical Bahcall--Wolf cusp for which $\gamma = 7/4$ \cite{BahcallWolf1976}. Observational constraints and theoretical studies, however, infer a broad range of inner slopes \cite{Alexander_1999,Genzel_2003,Schodel:2014NSCReview,Linial_2022}. Accordingly, rather than adopting a single best-fit value, we consider the physically motivated interval:
\begin{equation}
1 \lesssim \gamma \lesssim 2,
\end{equation}
thereby encapsulating the substantial uncertainties associated with the diversity of models and measurements.

Rather than normalizing the neutron star (NS) population of the nuclear star cluster (NSC) using the $M$–$\sigma$ relation, we instead adopt a normalization motivated by the star-formation history of the NSC \citep{Chen_2023,Wharton_2012}. Specifically, we normalize the cluster to its total neutron-star population which extends to a maximum of 200 pc. The predicted number of neutron stars depends sensitively on the assumed initial mass function (IMF). For example, \cite{Chen_2023} found that for Kroupa and top-heavy IMFs with varying metallicities, the number of neutron stars lies in the range $0.5$–$3.2 \times 10^4$ per $10^6,M_\odot$ of NSC mass. Likewise \cite{Wharton_2012}, found that by adopting either a standard Salpeter IMF or a top-heavy IMF yields approximately $\sim 4$–$6 \times 10^3$ neutron stars within $~1\,\rm pc$ and expanding the NS progenitor mass range can increase this estimate to $\sim 10^4$ which is consistent with \cite{Chen_2023, Flores:2023PRD} for different values of $\gamma$ \citep{Wharton_2012}.

Given these uncertainties, we adopt the representative range of $10^4$–$10^6$ neutron stars within the NSC. This choice allows for potentially enhanced Galactic Center densities while remaining compatible with galaxy-wide neutron star constraints. As we discuss in the next section, variations in this slope have important implications for the formation rate and spatial distribution of G objects.

\begin{table}[t]
\centering
\caption{Stellar mass components adopted for the Milky Way potential. Disk and bulge masses are taken from the analysis of \cite{LicquiaNewman2015, Rix_2013}. Nuclear Stellar Disk (NSD) and Nuclear Star Cluster (NSC) masses follow the nuclear-bulge decomposition summarized by \citet{Launhardt_2002} which used a similar density distribution.}
\label{tab:mw_masses}
\begin{tabular}{cccc}
\hline\hline
Disk & Bulge (+bar) & NSD & NSC \\
\hline
$5.2\times10^{10}\,M_\odot$ &
$9.1\times10^{9}\,M_\odot$ &
$1.4\times10^{9}\,M_\odot$ &
$3.0\times10^{7}\,M_\odot$ \\
\hline
\end{tabular}
\end{table}

\subsection{\label{sec:DM_distribution}Dark Matter Distribution}

To model the dark matter distribution we employ the commonly used two step power law \cite{Navarro_1996, Flores:2023PRD}:

\begin{equation}
    \rho_{\text{DM}} \simeq \frac{\rho_{\odot}}{(r/r_\odot)^{\alpha} (1+ r/r_\odot)^{\beta-\alpha}}
\end{equation}
where the exact scaling is set so the local dark matter density is $\rho_\odot = 0.4 \, \text{GeV  cm}^{-3}$\cite{McMillan_2016}
To calculate the velocity dispersion of the PBHs we assume a virialized halo such that \cite{Batista2011MOA2009BLG387Lb, DeRocco_2024}:

\begin{equation}
\label{v_rms}
    \sigma(r) = \sqrt{\frac{GM_{\text{tot}}(<r)}{r^2}}
\end{equation}

where  $M(<r)$ is the total enclosed mass given by $M_{\text{tot}} = M_{\text{SMBH}} + M_{\text{NSC}} + M_{\text{NSC}} + M_{\text{BLG}}  + M_{\text{DSK}}$  where the masses used are given in Tab.~\ref{tab:mw_masses}.

\subsection{PBH capture, settling, and accretion timescales}

In dark-matter-rich environments, a neutron star (NS) can be converted into a G objects through the capture of a PBH, followed by its subsequent settling and accretion inside the star. The total conversion timescale can be decomposed into three sequential components: 

(i) the PBH capture time, $t_{\rm cap}$; 

(ii) the settling time of the PBH inside the neutron-star core, $t_{\rm set}$; and 

(iii) the accretion time required for the PBH to consume the star and form a stellar-mass black hole, $t_{\rm acc}$. 

The total conversion timescale is therefore
\begin{equation}
t_{\rm con} \equiv t_{\rm cap} + t_{\rm set} + t_{\rm acc},
\end{equation}
and represents the time required for a neutron star to capture a PBH and evolve into a G objects.

The rate at which neutron stars capture primordial black holes (PBHs) is given by~\citep{Capela2013}

\begin{equation}
\Gamma_{\rm PBH}
= \frac{\Omega_{\rm PBH}}{\Omega_{\rm DM}}\,\Gamma^{\rm MW}_0
= f_{\rm DM}\,\Gamma^{\rm MW}_0 ,
\end{equation}

where $\Omega_{\rm PBH}$ and $\Omega_{\rm DM}$ denote the present-day energy-density fractions of PBHs and total dark matter, and $f_{\rm DM} \equiv \Omega_{\rm PBH}/\Omega_{\rm DM}$ is the fraction of DM that is in the form of PBHs. The quantity $\Gamma^{\rm MW}_0$ corresponds to the capture rate in the limiting case where PBHs constitute all of the dark matter. For a neutron star of mass $M_{\rm NS}$ and radius $R_{\rm NS}$, the reference capture rate is

\begin{equation} 
\begin{split} 
\Gamma^{\rm MW}_0 = \sqrt{6\pi}\, \frac{\rho_{\rm DM}}{M_{\rm PBH}\,\bar v} &\frac{2 G M_{\rm NS} R_{\rm NS}}{1 - 2 G M_{\rm NS}/R_{\rm NS}} \\& \times\left[ 1 - \exp\left(-\frac{3 E_{\rm loss}}{M_{\rm PBH}\,\bar v^{2}}\right) \right], 
\end{split} 
\end{equation}

where $\rho_{\rm DM}(r)$ is the local dark-matter density, and $\bar v \rightarrow \sigma $ is the characteristic PBH velocity dispersion.

The energy dissipated by a PBH during a single transit through the neutron star, due to dynamical friction and accretion, can be approximated as~\citep{Capela2013}

\begin{equation}
E_{\rm loss}
\simeq
58.8\,
\frac{G^{2} M_{\rm PBH}^{2} M_{\rm NS}}
{R_{\rm NS}^{2}} .
\end{equation}

Capture becomes efficient once the dissipated energy exceeds the initial kinetic energy of the PBH, $E_{\rm loss} \gtrsim \tfrac{1}{2} M_{\rm PBH} \bar v^{2}$. To account for the relative velocity between the neutron star and the PBH we assume the velocity distribution \cite{Caiozzo_2024}:,

\begin{equation}
f(v)
=
\sqrt{\frac{6}{\pi}}
\frac{v}{\sigma^{2}}
\sinh\!\left(\frac{3v}{\sigma}\right)
\exp\!\left[
-\frac{3\left(v^{2}+\sigma^{2}\right)}
{2 \sigma^{2}}
\right],
\end{equation}

The velocity-averaged capture rate is then

\begin{equation}
\left\langle \Gamma(r) \right\rangle
=
f_{\rm DM}
\int_0^\infty dv \,
f(v)\,
\Gamma^{\rm MW}_0(v),
\end{equation}

Finally, the corresponding capture timescale is

\begin{equation}
\left\langle t_{\rm cap} \right\rangle
=
\left\langle \Gamma(r) \right\rangle^{-1}.
\end{equation}

Once captured, the PBH undergoes repeated passages through the neutron star, losing orbital energy until it settles at the stellar center. The corresponding settling timescale is~\citep{Capela2013}
\begin{equation}
t_{\rm set} \simeq
1.3 \times 10^{9}\,{\rm yr}
\left( \frac{M_{\rm PBH}}{10^{19}\,{\rm g}} \right)^{-3/2}.
\end{equation}

After settling, the PBH accretes neutron-star material from the surrounding nuclear medium. We model this phase assuming spherical Bondi accretion, for which the PBH mass growth rate is~\citep{Bondi1952}
\begin{equation}
\dot M_{\rm PBH}
= 4\pi \lambda_s
\frac{G^2 M_{\rm PBH}^2 \rho_{\rm NS}}{v_s^{3}},
\end{equation} where $\rho_{\rm NS}$ is the characteristic neutron-star density, $v_s$ is the sound speed of nuclear matter, and $\lambda_s$ is an order-unity parameter that depends on the neutron-star equation of state. For a neutron star described by an $n=3$ polytropic equation of state, representative values are
$v_s \simeq 0.17\,c$, $\rho_{\rm NS} \simeq 10^{15}\,{\rm g\,cm^{-3}}$, and
$\lambda_s \simeq 0.707$~\citep{Flores:2023PRD}. The characteristic accretion
timescale is then
\begin{equation}
t_{\rm acc} \equiv \frac{M_{\rm PBH}}{\dot M_{\rm PBH}}
\simeq 10~{\rm yr}
\left( \frac{10^{19}\,{\rm g}}{M_{\rm PBH}} \right).
\end{equation}

\begin{figure}[t]
  \centering
  \includegraphics[width=0.95\columnwidth]{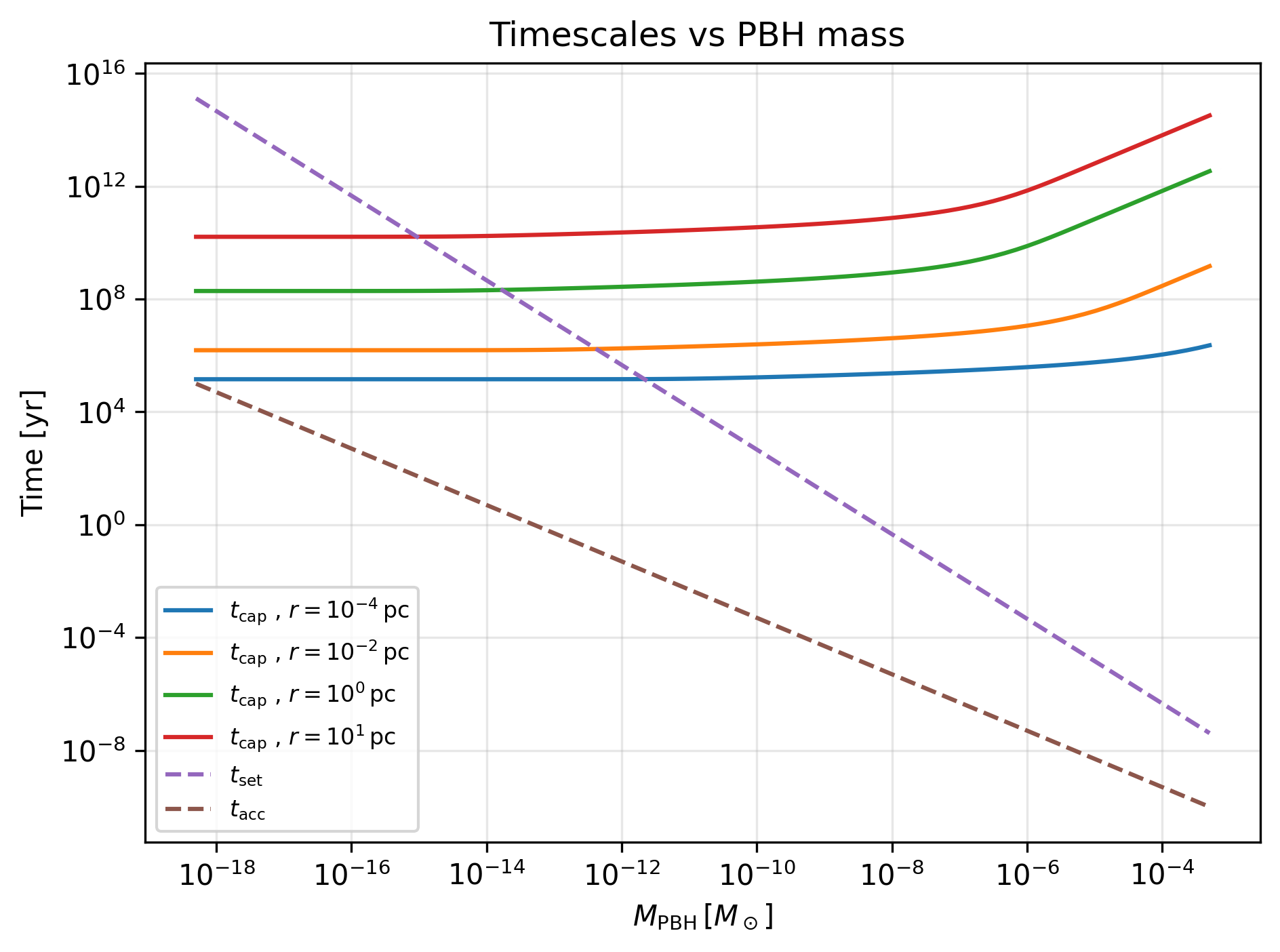}
\caption{Conversion timescales for PBH--neutron star encounters as a function of PBH mass,  for $\alpha = 2$, $f_{\rm DM} = 1$. At low masses the accretion timescales dominates,  while at high masses the capture rate increases due to the reduced number density of  heavier PBHs. Between these two regimes lies a valley where the total conversion rate is maximized; it is precisely in this mass window that PBHs must reside in order to reproduce  the observed population of G-object population.}
  \label{fig:conversion_timescales}
\end{figure}

Together, these three stages define the full neutron-star conversion process. As illustrated in Fig.~\ref{fig:conversion_timescales}, the conversion proceeds in three distinct regimes as a function of PBH mass: a low-mass regime in which the settling time dominates, an intermediate regime in which the capture rate dominates but is approximately mass-independent, and a high-mass regime in which the capture timescale scales linearly with $M_{\rm PBH}$.

%%%%%%%%%%%%%%%%%%%%%%%%%%%%%%%%%%%%%%%%%%%%%%%%%%%%%%%%%%%%

\subsection{Conversion probability and G objects number counts}
\label{sec: conversion}
To connect the conversion physics to observable populations, we model the
PBH-induced neutron-star conversion as a Poisson process. The probability that
a neutron star at Galactocentric radius $r$ has been converted over the age of
the Milky Way, $t_{\rm MW}$, is
\begin{equation}
\Upsilon(r) \equiv
1 - \exp\!\left[-\frac{t_{\rm MW}}{\langle t_{\rm con}(r) \rangle}\right],
\end{equation}
where $\langle t_{\rm conv}(r) \rangle$ is the conversion timescale averaged over the
local PBH velocity distribution.

The cumulative number of converted G objects within a spherical radius $r_0$
is then
\begin{equation}
N_G(r < r_0) =
\int_0^{r_0}
n_{\rm NS}(\vec r)\,
\Upsilon(r)\,
d^3 \vec r ,
\end{equation}
where $n_{\rm NS}(\vec r)$ is the neutron-star number density defined in Section \ref{sec:NSmodels}. Depending on the Galactic component considered, this density may be modeled using either cylindrical or spherical symmetry, as discussed in the preceding section.

\begin{figure}[t]
  \centering
  \includegraphics[width=0.95\columnwidth]{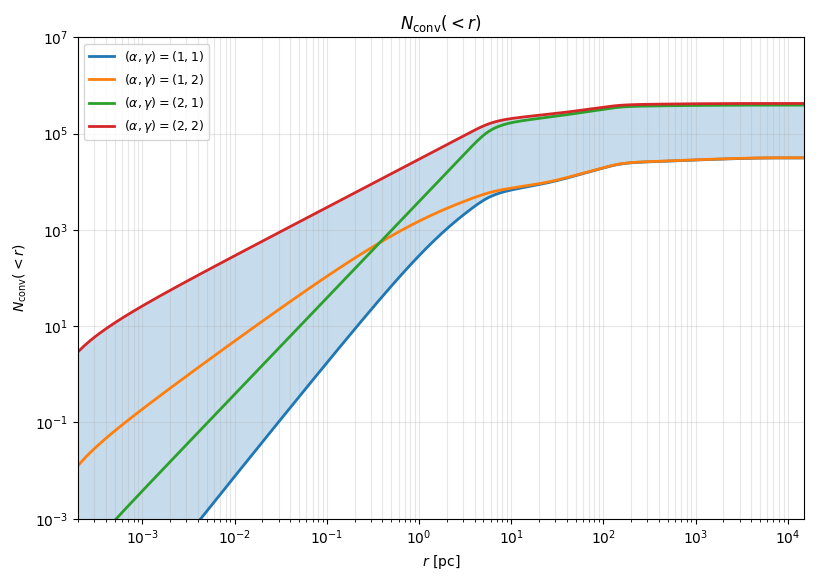}
\caption{Cumulative number of converted neutron stars within a spherical radius $r$ from the  Galactic centre, for $\alpha = 2$, $f_{\rm DM} = 1$, $M_{\rm PBH} = 10^{-9}\,M_\odot$.  Beyond $r \sim 100\,\mathrm{pc}$ the conversion rate drops sharply as both the PBH and  neutron star number densities fall off steeply, rendering G-object production negligible at larger Galactocentric distances. The observed G objects must therefore be confined to this inner region.}
  \label{fig:N_conv_MW}
\end{figure}
\subsection{Conversion probability and G objects number counts}
These expressions provide the bridge between PBH capture microphysics and the population-level predictions for G objects developed in the next section. Using the four component Galactic Milky Way (MW) model presented in the previous section we plot in Fig.~\ref{fig:N_conv_MW} the cumulative G objects count for a couple choice of model parameters showing the sensitivity on the inner dark matter and NS density slopes. We note that even for the most optimistic of models, i.e. $f_{DM} = 1,\,  N_{\text{NS,MW}} = 10^{9} $ the PBH to G objects conversion rate is strongly suppressed outside of about 100pc regardless of the choice of other parameters. This region corresponds to where the NSD dominates the density profile thus we should not expect to see any G objects in the Galactic bulge or disk. 

\begin{figure}[t]
  \centering
  \includegraphics[width=0.95\columnwidth]{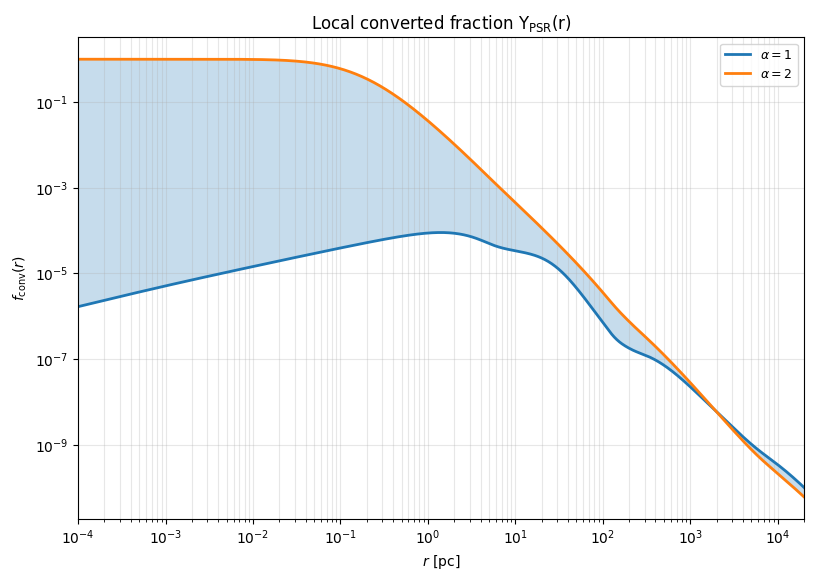}
   \caption{Radial dependence of the pulsar conversion fraction 
$\Upsilon_{\rm psr}(r) = 1 - \exp[-t_{\rm psr}/\langle t_{\rm con}(r)\rangle]$
for a representative PBH mass $M_{\rm PBH}=10^{-12}\,M_\odot$ and maximal dark-matter fraction $f_{\rm DM}=1$. 
Different curves correspond to varying inner dark-matter slopes $\alpha$. 
}
  \label{fig:frac_conv}
\end{figure}

Figure~\ref{fig:frac_conv} shows the radial dependence of the pulsar conversion fraction,
\[
\Upsilon_{\rm psr}(r)
=
1 - \exp\!\left[-\frac{t_{\rm psr}}{\langle t_{\rm con}(r)\rangle}\right],
\]
where $t_{\rm psr} \sim 10^7\,{\rm yr}$ \cite{FaucherGiguere2011} is the characteristic radio-loud lifetime of a pulsar. As the PBH capture rate scales with the local dark-matter density ($\Gamma \propto \rho_{\rm DM}$) and the inner halo behaves as $\rho_{\rm DM}\propto r^{-\alpha}$, efficient conversion requires sufficiently steep central density cusps.

For shallow or moderately cuspy profiles ($\alpha \lesssim 1.5$), the dark-matter density within the inner parsec is too low for capture to compete with the pulsar lifetime, and the conversion fraction remains negligible across the region accessible to current surveys. Even for steep cusps, only the extremal case $\alpha \simeq 2$ produces $\langle t_{\rm con}\rangle \lesssim t_{\rm psr}$, and then only within $\sim 0.1\,{\rm pc}$ of the Galactic Center. At larger radii, the conversion probability declines rapidly with $r$, tracking the underlying density profile.

Since most Galactic Center pulsar searches probe radii well beyond $1\,{\rm pc}$, PBH-induced conversion can at most deplete the innermost $\lesssim 0.1\,{\rm pc}$ and cannot account for the broader pulsar deficit. 
\subsection{Joint inference with the ``missing pulsar problem''}
\label{Missing_pulsar}
\paragraph{Galactic Center pulsar deficit and joint likelihood.}
Despite the inferred  high stellar density in the central parsec of the Galaxy, several deep radio surveys\cite{Bates_2010,Macquart_2010, Torne_2021}  have detected no normal pulsars in the Galactic Center (GC), with the sole exception of a single magnetar, PSR~J1745–2900, discovered in 2013 and located at a projected distance of $\sim 0.1$\,pc from Sgr~A* \cite{Mori:2013, Kennea_2013, Eatough2013Nature}.  This discrepancy between expectations and observations constitutes the so-called ``missing pulsar problem.

Many explanations have been proposed to address the so-called missing pulsar problem in the Galactic Center (GC). The discovery of PSR~J1745–2900 suggests that magnetar formation may be particularly efficient in the GC environment \cite{Dexter_2014}. Given the relatively short magnetar lifetimes of $\sim 10^4$~yr \cite{Beniamini_2019}, an enhanced magnetar birth rate could help explain the apparent deficit of ordinary radio pulsars. Magnetars are thought to originate from massive progenitors \cite{Figer2005, Muno2006}, which would be abundant under the proposed top-heavy initial mass function in the GC \cite{Lu2013}. They may also form from highly magnetized massive stars \cite{Ferrario2005} and through binary evolutionary channels \cite{Popov2006,Popov2009,Zhang2025}. Both scenarios are broadly consistent with the extreme conditions of the GC environment \cite{Dexter_2014}. Furthermore, population studies suggest that a substantial fraction, $\sim 0.1$--1, of neutron stars may be born as magnetars \cite{Beniamini_2019}, reinforcing the plausibility of efficient magnetar production in the GC.

Other studies \cite{MacquartKanekar2015, Rajwade:2017, Chennamangalam_2014} have argued that the Galactic Center pulsar population may be dominated by recycled millisecond pulsars  \cite{Pfahl2004,Wharton_2012,FaucherGiguere2011}. The stellar density within the central parsec ($\sim 10^6\,\mathrm{pc}^{-3}$; \cite{Genzel1996,Schoedel2007}) exceeds that of globular cluster cores by roughly two orders of magnitude, greatly enhancing the rate of close stellar encounters. In dense environments, the formation rate of low-mass X-ray binaries (LMXBs) scales approximately with the square of the stellar density \cite{Verbunt1987}, and LMXBs are well-established progenitors of recycled pulsars \cite{Alpar1982}. The observed overabundance of X-ray transients within $\sim 1\,\mathrm{pc}$ of Sgr~A* \cite{Muno2005} further supports the presence of dynamically formed compact binaries. Although ongoing star formation near the GC may produce young pulsars, the substantial neutron star population accumulated over previous star formation episodes, combined with enhanced dynamical interaction rates, makes it plausible that recycled MSPs dominate the present-day GC pulsar population. Due to their lower luminosities, studies \cite{Rajwade:2017,MacquartKanekar2015} have shown that null detections in surveys can still allow $\mathcal{O}(10^3)$ MSPs in the inner $1$ pc  depending on the degree of scattering. 

In this work, we construct a joint likelihood that combines the GC pulsar non-detections with the observed population of G objects, under the hypothesis that a fraction of neutron stars are converted by interactions with PBHs. The same underlying neutron-star population simultaneously determines the number of radio pulsars that survive PBH conversion and the number of converted objects contributing to the observed G objects sample.

\paragraph{G objects likelihood.}

Given the PBH-induced neutron-star conversion model defined in Sec.~\ref{sec: conversion}, the predicted population of converted neutron stars is fully specified by the cumulative number
\begin{equation}
N_{\rm conv}(<r).
\end{equation}
We treat the observed G objects as an inhomogeneous Poisson point process in radius, with reference intensity
\begin{equation}
\lambda_0(r) \equiv \frac{d N_{\rm conv}(<r)}{dr}.
\end{equation}
For a set of observed G-object radii $\{r_k\}$ within the analysis window $[r_{\min}, r_{\max}]$, the corresponding reference log-likelihood is
\begin{equation}
\ln \mathcal{L}_{G,0}
=
\sum_k \ln \lambda_0(r_k)
-
\int_{r_{\min}}^{r_{\max}} \lambda_0(r)\,dr .
\end{equation}

To account for incompleteness in the observed G-object sample, we introduce a thinning fraction, $\epsilon_G$, defined as the probability that a converted and observationally eligible neutron star is identified as a G object. This corresponds to a Poisson thinning of the underlying population.

For each G object, we assign a representative radius $r_k$ given by its time-averaged orbital separation, computed from the orbital parameters reported in~\cite{Ciurlo:2020Nature, Gillessen:2012G2, Gillessen_2026}. The analysis window is taken to be the OSIRIS field of view, a $3'' \times 2''$ region centered on Sgr~A*, corresponding to a projected spatial scale of approximately $0.07\,\mathrm{pc}$ at $R_0 = 7.971 \, \rm kpc.$.

Given this selection window, highly eccentric objects such as G5 spend a significant fraction of their orbit outside the field of view. Rather than modeling this geometrical selection explicitly, we absorb its effect into the effective detection probability $\epsilon_G$.

In addition, G objects can only form from isolated neutron stars. We therefore include a factor
\begin{equation}
f_{\rm iso} \equiv 1 - f_{\rm binary}\, f_{\rm surv},
\end{equation}
where $f_{\rm binary}$ is the birth binary fraction of massive stars in the Galactic Center, and $f_{\rm surv}$ is the fraction of binaries that survive natal kicks and remain bound after neutron-star formation. Studies of young stars have found that the fraction of binaries is $>0.42$\cite{Gautam2019BinaryGC} whereas another study found that $f_{\rm surv} \sim 11.8\%$\cite{Hoang2021BinaryKicksGC}. The factor $f_{\rm iso}$ thus represents the fraction of neutron stars that are isolated and capable of producing G objects.

Uncertainty in the overall neutron-star abundance is absorbed into a normalization factor $A$, defined by the enclosed number of neutron stars within the NSC. The observed intensity is therefore
\begin{equation}
\lambda_{\rm obs}(r)
=
\epsilon_G \, A \, f_{\rm iso} \, \lambda_0(r).
\end{equation}

Following Sec.~\ref{combined_incompleteness}, we place a uniform prior $\epsilon_G \in [0.01, 0.05]$. The resulting G-object log-likelihood becomes
\begin{equation}
\begin{split}
    \ln \mathcal{L}_{G}
=
\ln \mathcal{L}_{G,0}
+
N_{\rm obs}\,&\ln\!\left(\epsilon_G A f_{\rm iso}\right)\\&
-
\left(\epsilon_G A f_{\rm iso} - 1\right)
\int_{r_{\min}}^{r_{\max}} \lambda_0(r)\,dr ,
\end{split}
\end{equation}
where $N_{\rm obs}$ is the number of observed G objects in the analysis window.

In this formulation, the factors $\epsilon_G$, $A$, and $f_{\rm iso}$ modify only the overall expected number of detected objects, while leaving the predicted radial distribution unchanged..

\paragraph{Missing-pulsar likelihood (Monte Carlo survey selection).}
The absence of detected radio pulsars toward the Galactic Center (GC) provides a powerful constraint on any model that predicts a large population of radio-loud pulsars in the inner few parsecs. Rather than adopting an analytic detection efficiency based on an assumed luminosity function, we follow an approach in which the survey selection function is computed by Monte Carlo population synthesis and a realistic survey detection pipeline.

We generate synthetic populations of canonical pulsars (CPs) using the \texttt{PsrPopPy} framework \citep{Bates_2014}, drawing birth properties (spin period $P$, magnetic field strength, kick velocities, age, and spectral indices) from the same parametric distributions used in recent Galactic Center (GC) pulsar population studies \cite{Bates_2014, Rajwade:2017, FaucherGiguere:2006}. Each synthetic pulsar is evolved forward in time using the \texttt{PsrPopPy} evolution routines.  We then condition on pulsars that are both radio-active (i.e., have not crossed the radio death line) and beaming toward Earth. We fix the dispersion measure to ${\rm DM}=1778~{\rm pc\,cm^{-3}}$, consistent with the GC magnetar PSR~J1745$-$2900 \citep{Eatough_2013}. For each surviving pulsar, we assign a radio pseudo-luminosity following the empirical prescription adopted by \citet{Bates_2014},
parametrized as a power law in spin period and period derivative:
\begin{equation}
    L = \gamma\, P^{\alpha}\, \dot P^{\beta},
\end{equation}
where $\alpha$, $\beta$, and $\gamma$ are model parameters. 
The corresponding flux density at the GC distance $d_{\rm GC}$ is
\begin{equation}
    S = \frac{L}{d_{\rm GC}^2}.
\end{equation}

The resulting sample of $N$ ``alive+beaming'' pulsars is then passed through a simulated pulsar survey to determine whether each source would be detected by comparing it to the minimum detectable pseudo-luminosity
\begin{equation}
    L_{\rm min} = S_{\rm min}\, d_{\rm GC}^2,
\end{equation}
where $S_{\rm min}$ is given by the radiometer equation \citep{Bates_2014}:
\begin{equation}
S_{\rm min} =
\frac{(S/N)_{\rm min}\, T_{\rm sys}^{GC}}
     {G\, \sqrt{n_p\, t_{\rm int}\,\Delta\nu}}
\sqrt{\frac{W_{\rm eff}}{P - W_{\rm eff}}},
\end{equation}
where $(S/N)_{\rm min}$ is the adopted signal-to-noise detection threshold, $T_{\rm sys}^{GC}$ is the total system temperature toward the Galactic Center (including receiver and sky background contributions), $G$ is the telescope gain (K Jy$^{-1}$),  $n_p$ is the number of  polarizations,  $t_{\rm int}$ is the integration time, $\Delta\nu$ is the observing bandwidth, $P$ is the pulsar spin period, and $W_{\rm eff}$ is the effective pulse width. The effective pulse width is defined as
\begin{equation}
W_{\rm eff} =
\sqrt{W_{\rm int}^2 + \tau_s^2 + \Delta t_{\rm DM}^2 + \delta t^2},
\end{equation}
where $W_{\rm int} = \delta P$ is the intrinsic pulse width (with $\delta$ the intrinsic duty cycle), $\tau_s = 1.3 \nu_{\text{GHz}}^{-4}$ \cite{Spitler_2013} is the pulse broadening timescale due to multi-path scattering,  $\Delta t_{\rm DM}$ is the intra-channel dispersion smearing time, and $\delta t$ is the instrumental sampling time. A pulsar is classified as detected if $S \ge S_{\rm min}$ (equivalently $L \ge L_{\rm min}$).

\begin{table}[t]
\centering
\caption{Model parameters used in \texttt{PsrPopPy} for the simulation of canonical pulsars (CPs).}
\label{tab:cp_params}
\begin{tabular}{ll}
\hline
Parameter & CP \\
\hline

Luminosity parameter $\alpha$ & $-1.4$ \\
Luminosity parameter $\beta$ & $0.50$ \\
Luminosity parameter $\gamma$ & $0.35$ \\

Spectral index distribution & Gaussian \\
$\langle \alpha \rangle$ & $-1.4$ \\
$\sigma_\alpha$ & $0.9$ \\

Initial spin period distribution & Gaussian \\
$\langle P \rangle$ & $300\,\mathrm{ms}$ \\
$\sigma_P$ & $150\,\mathrm{ms}$ \\

Spin-down model & Faucher-Gigu\`ere \& Kaspi \cite{FaucherGiguere:2006} \\
Beam alignment model & Orthogonal \\
Braking index & $3.0$ \\

Initial $B$-field distribution & Log-normal \\
$\langle \log_{10} B(\mathrm{G}) \rangle$ & $12$ \\
$\sigma_{\log B}$ & $0.55$ \\
Survey  & \citet{Perez_2026} \\
Frequency(GHz) & 9.98\\
Bandwidth $\Delta \nu $(GHz) & 4.88\\

Scattering  & $\tau_{\text{WS}} = 1.3 \nu_{\text{GHz}}^{-4}$ \cite{Spitler_2013} \\
$S/N_{\min}$ & 3 \\
Gain & $2 \, \text{K J}^{-1}$ \\
integration time (hrs) & 2 \\
\hline
\end{tabular}
\end{table}

\begin{figure}[t]
  \centering
  \includegraphics[width=0.95\columnwidth]{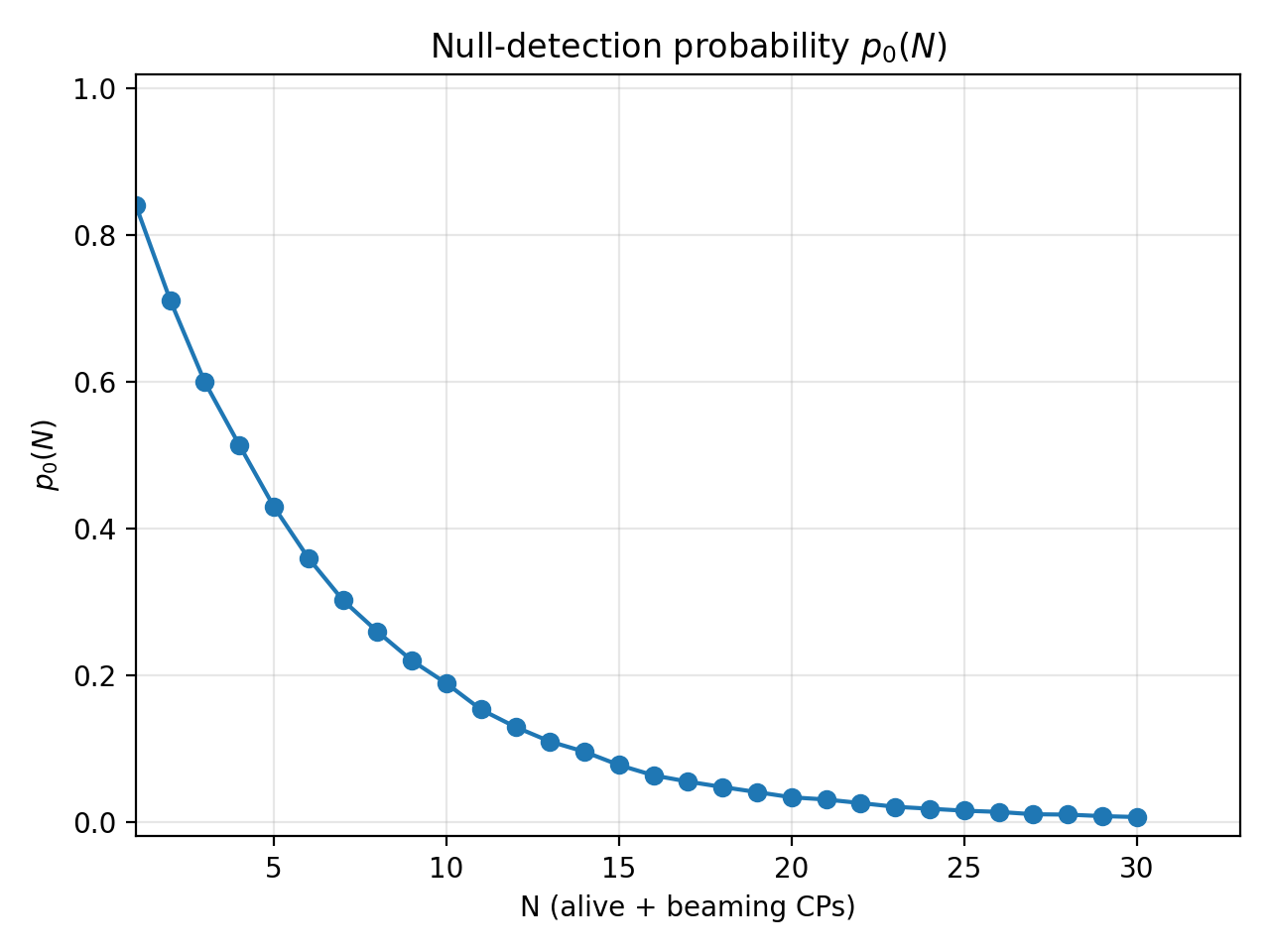}
\caption{Null-detection probability $p_0(N)$ as a function of the number of observable (alive and beaming) canonical pulsars $N$ in the Galactic Centre. Each point shows the fraction of Monte Carlo realizations  in which zero pulsars are detected by the survey generated using the population synthesis parameters of \citet{Rajwade:2017}.}
  \label{fig:p0_curve}
\end{figure}

In this work we adopt the most sensitive GC pulsar search to date: the Green Bank Telescope (GBT) X-band Galactic Center survey, whose full 2\,hr integrations reach a limiting flux density of \(S_{\min}\simeq 0.002~{\rm mJy}\) under the weak-scattering assumption \cite{Perez_2026}. We use the survey’s instrumental setup (central frequency and bandwidth) and its detection thresholding as implemented in the \texttt{PsrPopPy} survey simulator. Under this forward model, the probability of obtaining zero detections in the survey, given an underlying sample size \(N\) of alive+beaming pulsars in the searched region, is estimated as

\begin{equation}
p_0(N)
=
\frac{1}{N_{\rm trial}}
\sum_{t=1}^{N_{\rm trial}}
\mathbb{I}\!\left[k_t(N)=0\right],
\end{equation}
where \(k_t(N)\) is the number of detections in trial \(t\), and \(\mathbb{I}\) is the indicator function. Repeating this procedure over a grid \(N=0,1,\dots,N_{\max}\) yields a tabulated null-detection curve \(p_0(N)\) as shown in ~\ref{fig:p0_curve}. We then define an interpolating function \(\ln p_0(N)\) and use it directly as the likelihood contribution from the missing-pulsar constraint.

To connect the PBH conversion model to the expected alive+beaming pulsar count \(N\), we need to relate the number of neutron stars to the underlying CP population. Following \cite{Wharton_2012} and similar parameterization in \cite{Faucher_Gigu_re_2011, Lazio_Cordes_2008} we can relate the two quantities as: 

\begin{equation}
\label{N_CP}
    N_{\text{CP}} = f_{\text{PSR}}f_bf_\tau f_v N_{\text{NS}}
\end{equation}
Here $f_{\rm PSR} \sim 1$ denotes the fraction of neutron stars that become radio pulsars, and $f_b \simeq 0.1$ is the beaming fraction, i.e., the fraction of pulsars whose emission beams intersect our line of sight. 
These quantities are relatively well constrained by Galactic pulsar statistics. By contrast, $f_\tau$—the fraction of pulsars younger than the characteristic radio-loud lifetime—depends sensitively on the star-formation history of the Galactic Center. If star formation has proceeded continuously over long timescales, $f_\tau$ may be as low as $\sim 10^{-3}$ \citep{Wharton_2012}. However, studies indicating enhanced star formation $\sim 100\,{\rm Myr}$ ago \citep{Pfuhl_2011,Blum_2003} imply a substantially larger  fraction, $f_\tau \sim 0.1$ \citep{Wharton_2012}. Thus to not make any assumptions on the NSC star formation history we will take a log-uniform prior $[10^{-3}-10^{-1}]$. The neutron-star retention fraction in the Galactic Center, $f_v$, depends on the natal kick distribution and local escape velocity. 
Simple Maxwellian kick models predict $f_v \sim 0.1$ at $r \sim 1$ pc \citep{Wharton_2012}, though the exact value varies with the assumed velocity distribution. However, analogous calculations are known to underpredict retention in globular clusters, where dynamical interactions and high stellar densities substantially enhance neutron-star survival \citep{Pfahl_2002}. Given that the Galactic Center exhibits even more extreme densities and interaction rates, we adopt $f_v \sim 1$ as in \citep{Wharton_2012}. We interpret the parameter $f_p \equiv f_{\rm PSR} f_b f_\tau f_v$  as the effective fraction of neutron stars that would be radio-loud, younger than the radio-visible lifetime $T_{\rm vis}$, and beaming toward Earth in the absence of PBH-induced conversion. 

For model parameters $\boldsymbol{\theta}$, we compute the cumulative number of neutron stars that remain unconverted over the radio-visible lifetime, $T_{\rm vis} = 10^7$yrs, inside the survey radius $r_{\max}$ as
\begin{equation}
N_{\rm surv}^{(T_{\rm vis})}(<r_{\max}\,|\,\boldsymbol{\theta})
=
\int_0^{r_{\max}}
n_{\rm NS}(r\,|\,\boldsymbol{\theta})\,
S_{\rm psr}(r\,|\,\boldsymbol{\theta})\,
4\pi r^2\,dr,
\end{equation}
where $n_{\rm NS}(r)$ is the neutron-star number density and
\begin{equation}
S_{\rm psr}(r\,|\,\boldsymbol{\theta})
=
\exp\!\left[-\frac{T_{\rm vis}}{\tau(r\,|\,\boldsymbol{\theta})}\right]
\end{equation}
is the survival probability for remaining unconverted over $T_{\rm vis}$, expressed in terms of the effective PBH conversion timescale $\tau(r)$.

The predicted number of alive+beaming pulsars in the survey region is then
\begin{equation}
N_{\rm alive+beam}(\boldsymbol{\theta})
=
f_p\,
N_{\rm surv}^{(T_{\rm vis})}(<r_{\max}\,|\,\boldsymbol{\theta}),
\end{equation}
and the missing-pulsar likelihood contribution is
\begin{equation}
\ln \mathcal{L}_{\rm PSR}(\boldsymbol{\theta})
=
\ln p_0\!\left(N_{\rm alive+beam}(\boldsymbol{\theta})\right),
\end{equation}
where $\ln p_0$ is obtained by interpolation of the Monte Carlo null-detection curve. 

This construction incorporates the full survey selection function through the Monte Carlo calibration of $p_0(N)$, while allowing PBH-induced conversion to deplete the radio-visible pulsar population within the finite lifetime window $T_{\rm vis}$.
\paragraph{Joint inference.}
Both likelihood components depend on the same underlying neutron-star population and PBH conversion physics. The G-object likelihood constrains the cumulative number and radial distribution of neutron stars converted over the Galactic age, while the pulsar likelihood constrains the complementary population that remains unconverted over the radio-visible lifetime. The two data sets therefore probe complementary phases of the same neutron-star population within a single PBH conversion framework.

We perform a fully joint Bayesian inference over the physical parameters of interest and all relevant nuisance parameters. Let $\bm{\theta}$ denote the PBH and density-profile parameters,
\[
\bm{\theta} = \{\alpha_{\rm DM},\,\gamma,\,M_{\rm PBH},\,f_{\rm DM}\},
\]
and let $\bm{\phi}$ denote the astrophysical and survey nuisance parameters,
\[
\bm{\phi} = \{f_p,\, f_{\rm iso}, \,N_{\rm NS,NSC},\,\epsilon_G\}.
\]
The full posterior is then
\begin{equation}
p(\bm{\theta},\bm{\phi}\mid\mathcal{D})
\;\propto\;
\mathcal{L}_{G}(\bm{\theta},\bm{\phi})\,
\mathcal{L}_{\rm PSR}(\bm{\theta},\bm{\phi})\,
p(\bm{\theta})\,p(\bm{\phi}),
\end{equation}
where $\mathcal{D}$ represents the combined G-object and radio data.

Rather than performing analytic marginalization, we sample the joint posterior distribution using a Markov Chain Monte Carlo (MCMC) method. Posterior constraints on the physical parameters $\bm{\theta}$ are obtained by numerical marginalization over the nuisance parameters $\bm{\phi}$ through the MCMC samples,
\[
p(\bm{\theta}\mid\mathcal{D})
=
\int
p(\bm{\theta},\bm{\phi}\mid\mathcal{D})\,
d\bm{\phi}.
\]

To avoid imposing informative assumptions, we adopt broad priors summarized in Table~\ref{tab:priors}. The neutron-star cusp slope is taken in the range $\gamma\in[1,2]$, and the inner dark-matter slope in $\alpha_{\rm DM}\in[0,2]$. The PBH mass is assigned a log-uniform prior over $M_{\rm PBH}\in[10^{14},10^{33}]\,{\rm g}$, and the PBH dark-matter fraction a log-uniform prior over $f_{\rm DM}\in[10^{-10},1]$. Regions of $(M_{\rm PBH},f_{\rm DM})$ parameter space excluded by evaporation and microlensing constraints are removed via hard prior cuts following Ref.~\cite{Carr_2021}. 

\begin{figure*}[t]
  \centering
  \includegraphics[height=0.55\textheight]{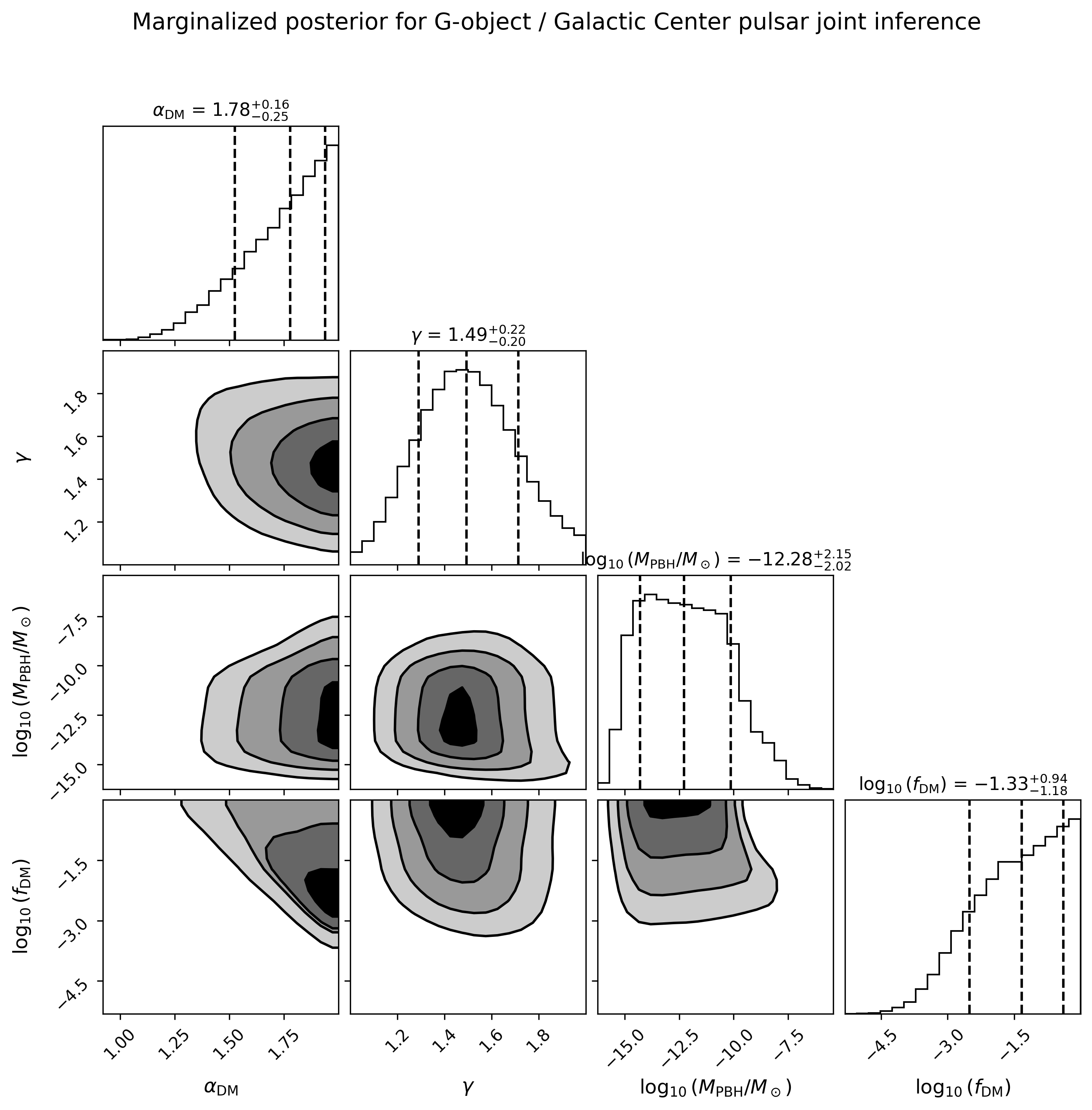}
  \caption{
Marginalized Bayesian posterior distributions for the physical parameters of the model, obtained by jointly sampling the PBH and astrophysical parameters together with all survey and normalization nuisance parameters using MCMC. Shown are the one- and two-dimensional marginalized posteriors for the dark matter inner slope $\alpha$, the neutron star cusp slope $\gamma$, the PBH mass $M_{\rm PBH}$, and the PBH dark matter fraction $f_{\rm DM}$. Contours correspond to the 68\% and 95\% credible regions, while vertical lines indicate the median and central credible intervals. All nuisance parameters, including the neutron star normalization and survey completeness, have been fully marginalized over. The posterior exhibits a preference for cuspy density profiles as well as broad $m_{\rm PBH}$ peak reflecting the the current PBH abundance constraints specifically those of HSC \cite{sugiyama2026}.
}

  \label{fig:Bayesian}
\end{figure*}

The nuisance parameters $f_p$, $N_{\rm NS,NSC}$, and $\epsilon_G$ are assigned broad priors reflecting uncertainties in pulsar demographics, neutron-star normalization, and G-object completeness, respectively. All nuisance parameters are sampled jointly with the physical parameters and marginalized over when presenting posterior constraints in Fig.~\ref{fig:Bayesian}.

In this formulation, the posterior explicitly encodes the requirement that a single PBH model simultaneously reproduce (i) the observed G-object distribution, (ii) the absence of detected canonical pulsars within the survey region, and (iii) external PBH abundance constraints.

\begin{table}[h]
\centering
\begin{tabular}{ll}
\hline
Parameter & Prior \\ 
\hline
$\alpha_{\rm DM}$ 
    & Uniform on $[0,\,2]$ \\

$\gamma$ 
    & Uniform on $[1,\,2]$ \\

$M_{\rm PBH}$ 
    & Log-uniform on $M_g \in [10^{14},\,10^{33}]\,{\rm g}$ \\

$f_{\rm DM}$ 
    & Log-uniform on $[10^{-10},\,1]$, as well as \\ 
    & evaporation/microlensing constraints\\

$f_p$ 
    & Log-Uniform on $[10^{-4},\,2\times10^{-2}]$ \\

$N_{\rm NS,NSC}$ 
    & Uniform on $[10^{5},\,10^{6}]$ \\

$\epsilon_G$ 
    & Uniform on $[0.01,\,0.05]$ \\
    
$f_p$ 
    & Log-Uniform on $[10^{-4}, \, 10^{-2}]$ \\
\hline
\end{tabular}
\caption{Priors used in the joint PBH–Galactic Center inference. Parameters sampled uniformly in $\log_{10}$ correspond to log-uniform priors in physical space.}
\label{tab:priors}
\end{table}
\indent The posterior distribution for the PBH mass spans a broad range, with 
\[
\log_{10}\!\left(M_{\rm PBH}/M_\odot\right) = -12.28^{+2.15}_{-2.02},
\]
reflecting the weak dependence of the neutron-star (NS) lifetime on $m_{\rm PBH}$, as shown in Fig.~\ref{fig:conversion_timescales}. The posterior sharply declines below $\sim 10^{-15}\,\Msun$, where the settling timescale becomes so long that the NS lifetime exceeds the age of the Milky Way; this region is also excluded by PBH evaporation constraints \cite{Carr_2021}.  

\indent Toward higher masses, the posterior decreases more gradually, tracing the maximum allowed dark matter fraction, $f_{\rm DM}$, at fixed $m_{\rm PBH}$ under the updated HSC microlensing constraints \cite{sugiyama2026}.  

\indent The inferred dark matter fraction,
\[
\log_{10} f_{\rm DM} = -1.33^{+0.94}_{-1.33},
\]
corresponds to a broad region of parameter space in which PBHs may constitute a substantial fraction of the dark matter. Within the microlensing-allowed mass windows, the presence of the current G-object population is therefore consistent with scenarios in which PBHs account for an appreciable, and potentially dominant, component of the dark matter.
In contrast, the marginalized posteriors show a marked preference for centrally concentrated density profiles. The inner dark matter slope is constrained to $\alpha = 1.78^{+0.16}_{-0.25}$, indicating a cuspy halo in the Galactic Center. Likewise, the neutron-star density profile favors a moderately steep cusp with $\gamma = 1.49^{+0.22}_{-0.20}$, in agreement with stellar density measurements modeled using Nuker profiles$\gamma = 1.41 \pm 0.06 \pm \rm \, 0.1sys $\citep{Gallego_Cano_2017,Lauer_2007}. Physically, these steeper profiles increase the compact-object interaction rate in the inner parsec, thereby enhancing PBH capture and neutron-star conversion where the G-object population is observed, while naturally suppressing excessive production at larger radii.

\begin{figure*}[t]
  \centering
  \subfloat[$N_{\rm G}(<r)$\label{fig:nconv}]{
    \includegraphics[width=0.45\textwidth]{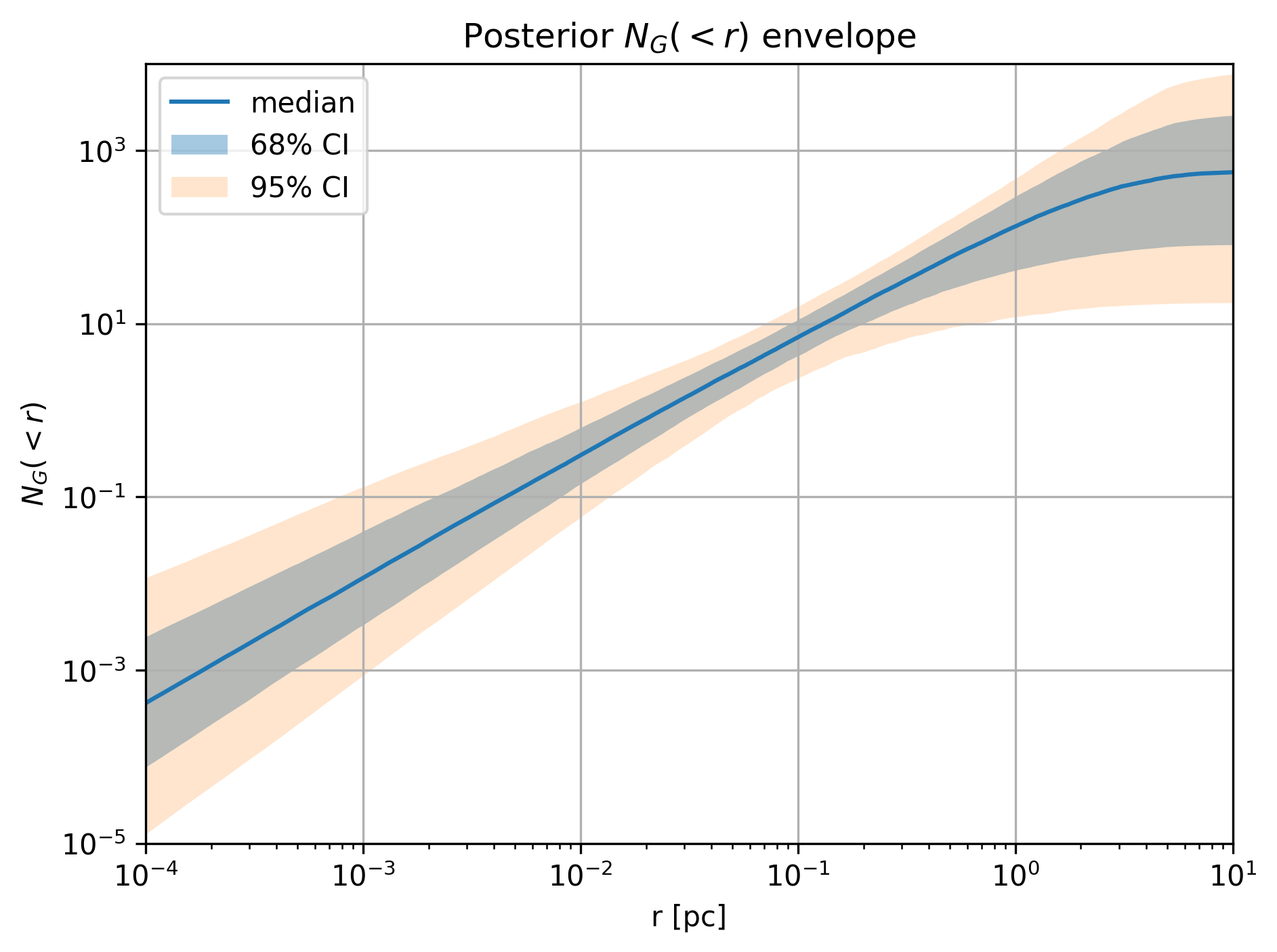}
  }\hfill
  \subfloat[$N_{\rm PSR}(<r)$\label{fig:npsr}]{
    \includegraphics[width=0.45\textwidth]{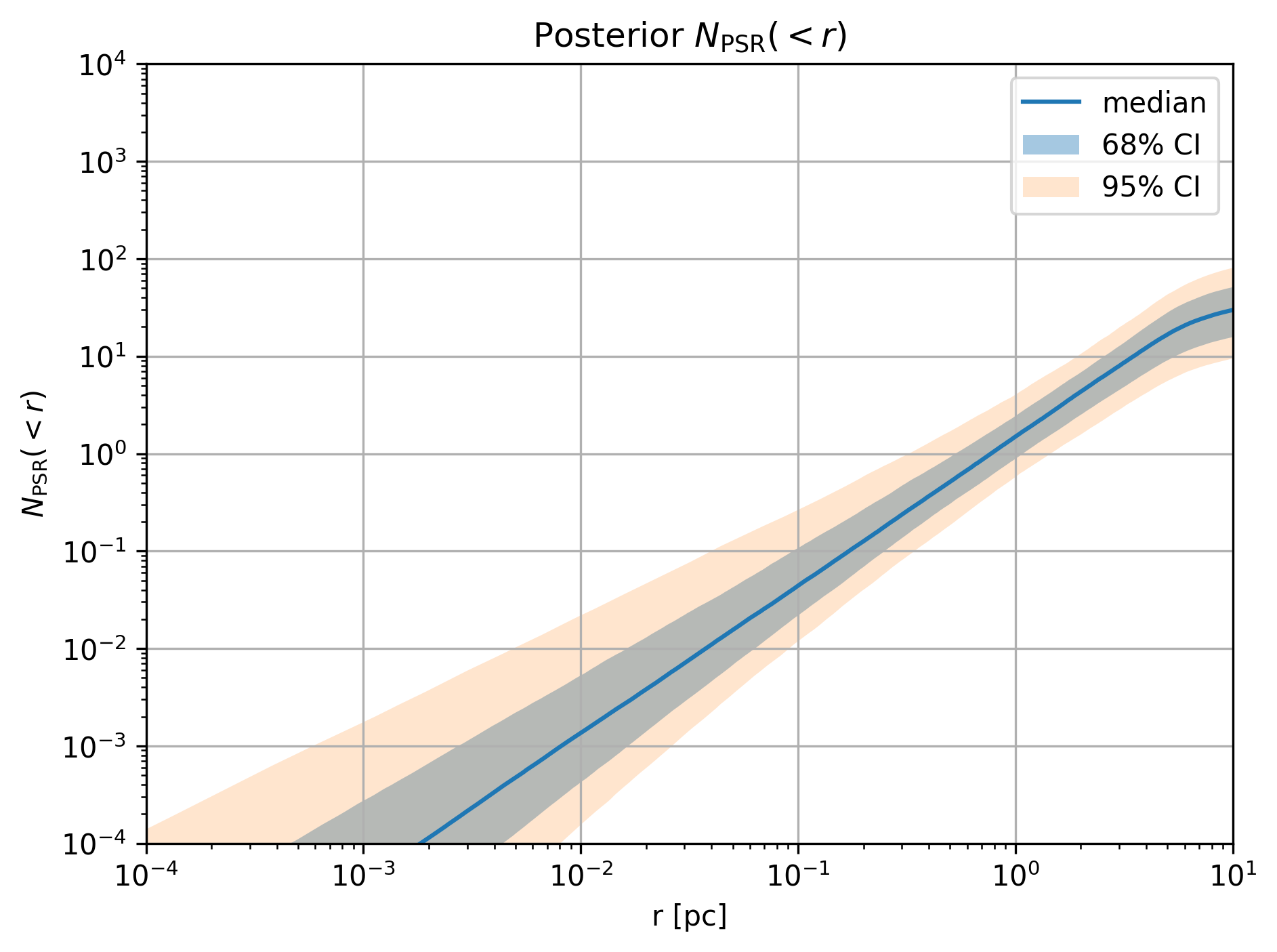}
  }

  \caption{Posterior envelopes for cumulative number of G objects $N_{\rm G}(<r)$,  and number of pulsars beaming towards us $N_{\rm PSR}(<r)=f_p\,\,N_{\rm surv}(<r)$ with $f_{\rm beam}=0.1$. Solid lines show the posterior median; shaded bands show the 68\% and 95\% credible intervals.}
  \label{fig:envelopes_all}
\end{figure*}

Figure~\ref{fig:envelopes_all} presents the posterior predictive distributions for the cumulative number of converted remnants, $N_{\rm G}$,  and the number of radio pulsars beaming toward Earth, $N_{\rm psr}$. Within the preferred parameter region, the model simultaneously reproduces the observed G-object counts and remains compatible with the absence of detected canonical pulsars. Importantly, the inference does not require large-scale pulsar destruction. Instead, the data favor a scenario in which the conversion rate over the radio-visible lifetime is negligible, implying that $\sim 0$ pulsars are expected to be destroyed. The missing-pulsar constraint therefore acts primarily as a prior on the underlying neutron-star population through Eq.~\ref{N_CP}, shaping the allowed ranges of $f_p$, $N_{\rm NS,NSC}$, and $\gamma$ rather than demanding substantial PBH-induced depletion.

%%%%%%%%%%%%%%%%%%%%%%%%%%%%%%%%%%%%%%%%%%%%%%%%%%%%%%%%%%
\subsection{Millisecond Pulsar Binaries and PBH Capture in the Galactic Center}
\label{sec:MSP_PBH_unified}

An alternative compact-object interpretation of G objects is that their central cores are millisecond pulsars (MSPs) in binaries, possibly embedded in dusty envelopes that account for the observed infrared continuum and recombination-line emission. This scenario raises two distinct but related questions: (i) whether MSP binaries are dynamically and radiatively compatible with G-object phenomenology, and (ii) whether PBH capture in MSP binaries is enhanced or suppressed relative to isolated neutron stars.

\paragraph{Tidal stability and internal perturbations.}
For a compact core of mass $m_{\rm core}$ on a highly eccentric orbit about Sgr~A$^*$, the Hill radius at pericenter $r_p$ is
\begin{equation}
r_H \simeq r_p \left( \frac{m_{\rm core}}{3M_\bullet} \right)^{1/3}.
\end{equation}
Adopting $M_\bullet \simeq 4\times10^6\,M_\odot$, $m_{\rm core} \sim 1$--$2\,M_\odot$, and $r_p \sim \mathcal{O}(10^2)$ AU as inferred for G2-like objects, one finds $r_H \sim \mathcal{O}(1)$ AU. Any envelope bound to the core must therefore remain confined within AU scales at pericenter. 

Typical MSP binaries have separations $a_{\rm bin} \sim 10^{-2}$--$10^{-1}$ AU, satisfying $a_{\rm bin} \ll r_H$. Thus, the Galactic-center tidal field does not automatically disrupt an MSP binary core. However, the presence of a close companion introduces distinctive signatures absent in single-core models. At 8 kpc, 1 AU subtends $\sim 125\,\mu{\rm as}$; a binary with $a_{\rm bin} \sim 0.1$ AU implies centroid excursions of order $\sim 10\,\mu{\rm as}$, potentially detectable with precision interferometry. Similarly, binary orbital velocities (hundreds of km\,s$^{-1}$) could induce periodic modulation of Br$\gamma$ line centroids on timescales of hours to days. The absence of short-period astrometric or spectroscopic modulation would therefore constrain an MSP-binary origin.

\paragraph{Envelope survival and radiative environment.}
A more significant challenge concerns energetics. Recycled MSPs launch relativistic winds powered by spin-down luminosity. In interacting systems, these winds ablate companions and drive intrabinary shocks. Sustaining a dusty AU-scale envelope requires dust survival against pulsar-wind irradiation and shock heating. Generically, one expects enhanced ionization and nonthermal radio/X-ray emission if an active MSP is present. By contrast, stellar-core models attribute the infrared luminosity primarily to reprocessed starlight with modest tidal heating, producing a comparatively quiescent spectral energy distribution. Moreover, MSPs are typically old ($\gtrsim$ Gyr) systems formed via binary recycling and do not naturally possess dense dusty envelopes, implying that an additional recent mass-ejection event would be required. 

\paragraph{PBH capture in MSP binaries versus isolated neutron stars.}
If G objects arise from PBH-induced neutron-star destruction, the capture rate is sensitive to the relative velocity between the neutron star and the PBH population. For isolated neutron stars, the capture rate scales approximately as
\begin{equation}
\Gamma_{\rm iso} \propto v_\infty^{-1}
\end{equation}
(or more steeply, depending on the energy-loss criterion). For an MSP in a binary with systemic velocity $v_{\rm sys}$ and orbital velocity $v_{\rm orb}$, the effective encounter velocity is
\begin{equation}
v_{\rm eff}^2 \simeq v_{\rm sys}^2 + v_{\rm orb}^2.
\end{equation}
In the Galactic Center, $v_{\rm sys} \sim 200$--$400$ km\,s$^{-1}$ and $v_{\rm orb} \sim 100$--$500$ km\,s$^{-1}$, so binary motion increases $v_{\rm eff}$. To leading order,
\begin{equation}
\frac{\Gamma_{\rm bin}}{\Gamma_{\rm iso}}
\simeq
\left(
\frac{v_{\rm sys}}{\sqrt{v_{\rm sys}^2+v_{\rm orb}^2}}
\right)^{\alpha},
\end{equation}
with $\alpha \gtrsim 1$. For $v_{\rm orb} \sim v_{\rm sys}$, one obtains $\Gamma_{\rm bin}/\Gamma_{\rm iso} \sim 2^{-\alpha/2}$, implying a suppression by a factor of a few for $\alpha \sim 1$--$2$. 

Unlike stellar three-body encounters, PBH capture requires dissipative energy loss inside the neutron star. The binary companion does not introduce an efficient additional dissipative channel, so MSP binaries are not expected to exhibit enhanced capture rates relative to isolated neutron stars. Instead, $\Gamma_{\rm bin} \lesssim \Gamma_{\rm iso}$.

\paragraph{Implications.}
These considerations lead to two conclusions. First, if a G object were unambiguously identified as the remnant of an MSP destroyed by PBH capture (e.g., via evidence for a compact $\sim$NS-mass black hole with MSP ancestry), this would provide especially strong support for the PBH-capture hypothesis, since capture in binaries is not expected to be enhanced. Second, the absence of MSP-origin remnants does not disfavor PBH-induced conversion: the intrinsic velocity suppression naturally predicts that PBH destruction should preferentially affect isolated neutron stars rather than compact MSP binaries.

In summary, while MSP binaries are not dynamically excluded as G-object progenitors, the combined constraints from envelope survival, binary-induced variability, and suppressed PBH capture rates render them less natural contributors than isolated neutron stars within the PBH-conversion framework.

%\newpage
%\clearpage

\section{Multi-wavelength observables and search strategies}\label{sec:multiw}
With guidance on the population of G objects and on the features of the  underlying population of PBHs, we now move on to a discussion of diagnostics for this scenario, and contrast the latter with standard G objects structures.

%\spr{REVIEWED UP TO HERE, JAN 29}

Upon capture in a neutron star (NS), PBHs settle to the stellar center and accrete the stellar material from the inside. The resulting contraction of the NS reduces its moment of inertia, and conservation of angular momentum drives a spin-up of the star. Because the contraction is more pronounced in the inner regions than in the outer layers, differential rotation could in principle develop. However, angular momentum transport via viscosity and magnetic stresses operates efficiently on timescales short compared to the PBH accretion time, allowing the star to maintain near rigid-body rotation. Continued spin-up increases the equatorial velocity, and once the latter exceeds the local escape velocity, matter is centrifugally ejected from the star. The expelled material can be substantial; for maximally rotating neutron stars (i.e., millisecond pulsars), the ejected mass can reach $\sim 0.1\,\Msun$ \cite{Fuller_2017,Takhistov_2019}. However, as discussed above, only isolated neutron stars can form G-objects, implying that the relevant spin periods are expected to be significantly longer than those of millisecond pulsars. Our posterior inference further indicates that the progenitors are radio-dead neutron stars that have long since crossed the pulsar death line. The spin period must therefore be evolved to the time at which the neutron star is converted (i.e., destroyed) via PBH capture. Consequently, the relevant time variable in the spin-down evolution is not an arbitrary stellar age, but the neutron star lifetime evaluated at the destruction time.
\subsection{Ejected Mass}
At fixed Galactocentric radius $r$, we model the PBH-induced conversion time as an exponential waiting-time distribution with mean $\tau(r)$. The corresponding cumulative distribution is
\begin{equation}
F(t)=1-e^{-t/\tau(r)}.
\end{equation}
Since we only consider NSs that convert within the Milky Way age $T_{\rm MW}$, we condition on the event $T\le T_{\rm MW}$. The conditional cumulative distribution is therefore
\begin{equation}
F(t\,|\,t<T_{\rm MW})
=\frac{1-e^{-t/\tau(r)}}{1-e^{-T_{\rm MW}/\tau(r)}}.
\end{equation}
Defining the median destruction time $t_{50}(r)$ via 
$F(t_{50}\,|\,t<T_{\rm MW})=1/2$, we obtain
\begin{equation}
t_{50}(r)
=\tau(r)\,\ln\!\left[
\frac{2}{1+\exp\!\left(-T_{\rm MW}/\tau(r)\right)}
\right].
\end{equation}
This quantity represents the characteristic NS lifetime at the moment of destruction.

\begin{figure}[t]
\centering
\includegraphics[width=0.85\linewidth]{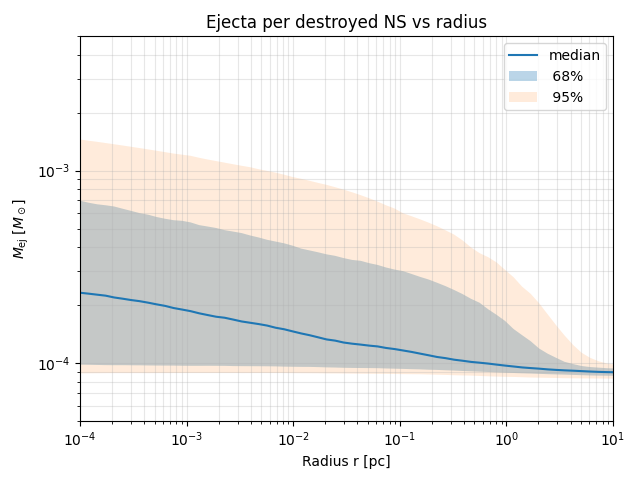}
\caption{
Expected ejecta mass per destroyed neutron star as a function of Galactocentric radius. For each posterior draw we compute the effective destruction timescale $\tau(r)$ and evaluate the conditional median destruction time $t_{50}(r)$ assuming an exponential waiting-time distribution truncated at $T_{\rm MW}=1.3\times10^{10}\,$yr.
The spin period at destruction is obtained from magnetic-dipole spin-down ($n=3$), and $M_{\rm ej}$ is evaluated using the polytropic ejection model.
Curves show the posterior median with 68\% and 95\% credible intervals for $n=3.0$.
}
\label{fig:Mej_vs_radius}
\end{figure}

We then evolve the spin period to $t=t_{50}(r)$ assuming standard magnetic–dipole spin-down with braking index $n=3$ \cite{Ridley_2010},
\begin{equation}
P(t)=\sqrt{P_0^{\,2}+2kB^2 t},
\end{equation}
where $P_0$ is the birth period and $k=9.77\times10^{-40}$ (cgs) for canonical neutron-star parameters ($R=10^6\,\mathrm{cm}$, $I=10^{45}\,\mathrm{g\,cm^2}$), with $B$ in Gauss and $P$ and $t$ in seconds. To compute $M_{\rm ej}$ we follow \cite{Fuller_2017}, adopting an $n=3$ polytropic model consistent with our earlier $t_{\rm acc}$ calculation. Because the neutron-star age at the time of capture depends sensitively on the PBH capture rate—and therefore on Galactocentric radius—we evaluate the ejected mass as a function of radius (Fig.~\ref{fig:Mej_vs_radius}). We find that neutron stars in the Galactic Center can eject between $\sim 10^{-4}$--$10^{-3}\,\Msun$, with a fraction of this material expected to remain gravitationally bound, s providing the observational basis for the G-objects as originally proposed in \cite{Flores:2023PRD}.

\subsection{BlackBody Fits}

It has been shown by \citet{Peissker2020} that external irradiation by the nearby S-star population is insufficient to account for the elevated dust temperatures inferred for the D-sources. For a dusty clump exposed to a UV radiation field, the equilibrium temperature can be approximated as \citep{Barvainis1987}
\begin{equation}
T_{\rm dust} =
9.627
\left[
\frac{L_{\rm UV}}{L_\odot}
\left(\frac{1\,{\rm pc}}{r_d}\right)^2
e^{-\tau_{\rm UV}}
\right]^{1/5.6}
\,{\rm K},
\end{equation}
where $L_{\rm UV}$ is the UV luminosity of the heating source, $r_d$ is the dust distance, and $\tau_{\rm UV}$ is the optical depth. Applying this relation to the radiation field of the S-star cluster, \citet{Peissker2020} derive equilibrium dust temperatures of only $T \sim 200$--$300\,\mathrm{K}$, comparable to other dusty filaments in the central parsec but well below the temperatures inferred for the D-sources.
\begin{figure}[ht]
    \centering
    \includegraphics[width=0.8\linewidth]{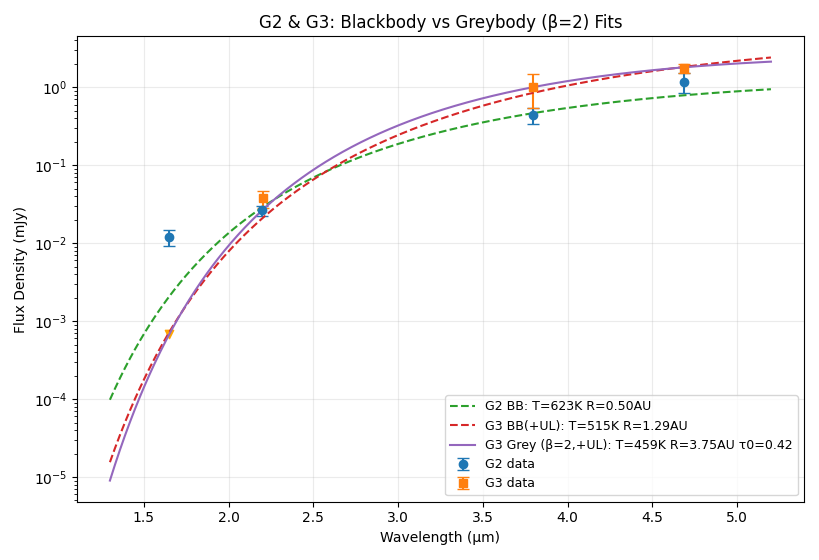}
    \caption{
    Infrared SEDs of G2 and G3 with best-fit blackbody and greybody models. 
    Photometric measurements are compiled from \citet{Peissker2020}, 
    \citet{Gillessen2012}, and \citet{Eckart_2013}. 
    The H-band upper limit for G3 is indicated by a downward triangle. A greybody with a larger radius is statistically favored by the current infrared emissions of G3. 
    }
    \label{fig:G2_G3_BB}
\end{figure}
To determine the characteristic dust temperatures of G2 and G3, we fit their infrared spectral energy distributions using both a pure blackbody model and a modified blackbody (greybody) model. The photometric measurements are taken from \cite{Peissker2020, Gillessen2012, Eckart_2013}. For the blackbody model we assume optically thick emission such that
\[
F_\nu = \left(\frac{R}{d}\right)^2 \pi B_\nu(T),
\]
where $R$ is the emitting radius, $d$ is the distance to the Galactic Centre, and $B_\nu(T)$ is the Planck function. For the greybody model we relax the assumption of complete optical thickness and adopt
\[
F_\nu = \left(\frac{R}{d}\right)^2 \pi B_\nu(T)\left(1 - e^{-\tau_\lambda}\right),
\]
with $\tau_\lambda = \tau_0 (\lambda_0 / \lambda)^\beta$, fixing $\beta=2$ which is typical for interstellar type grains \cite{Draine_2003} and defining $\tau_0$ at $\lambda_0 = 3.8\,\mu{\rm m}$. The results for the fits are shown in Fig \ref{fig:G2_G3_BB}.

For G2, fitting only the detected bands (four photometric points) yields $\chi^2 = 15.9$ for the blackbody model with $N_{\rm dof} = 2$ degrees of freedom (two fitted parameters: $T$ and $R$), corresponding to $\chi^2_{\rm red} = 7.97$ and best-fit values $T = 623\,{\rm K}$ and $R = 0.50\,{\rm AU}$. The greybody fit (three free parameters: $T$, $R$, $\tau_0$) converges to $\tau_0 = 52.5$, i.e. effectively optically thick emission, and yields the same $\chi^2 = 15.9$ but with $N_{\rm dof} = 1$. Since the additional parameter does not improve the goodness-of-fit, the blackbody model provides the statistically preferred description of G2.

For G3, three photometric detections (K, L, and M bands) are supplemented by an H-band non-detection. The non-detection is incorporated via a censored (one-sided) Gaussian likelihood. The detection-only blackbody fit yields $\chi^2 = 3.55$ with $N_{\rm dof} = 1$. For the full fit including the upper limit, we minimize the negative log-likelihood (NLL), defined as $\mathrm{NLL} = -\ln \mathcal{L}$, where $\mathcal{L}$ includes Gaussian terms for the detections and a cumulative Gaussian term for the upper limit. The blackbody model gives $\mathrm{NLL} = 2.57$, while the greybody model yields $\mathrm{NLL} = 1.36$.

To assess whether the additional greybody parameter is warranted, we compute the Akaike Information Criterion (AIC) and Bayesian Information Criterion (BIC), defined as
\[
\mathrm{AIC} = 2k + 2\,\mathrm{NLL},
\qquad
\mathrm{BIC} = k \ln(n) + 2\,\mathrm{NLL},
\]
where $k$ is the number of fitted parameters and $n$ is the number of data points. For G3 (with $n=4$ data constraints: three detections and one upper limit), the blackbody model yields ${\rm AIC}=9.14$ and ${\rm BIC}=7.91$, while the greybody model yields ${\rm AIC}=8.72$ and ${\rm BIC}=6.87$. The lower AIC and BIC values indicate that the improvement in fit quality is not solely due to the additional parameter, and that the greybody model is modestly preferred for G3. The resulting characteristic temperatures are $T \gtrsim 450\,{\rm K}$ for both sources, significantly higher than expected from external irradiation alone and therefore suggestive of an internal heating mechanism.

If the heating source were a compact remnant of mass $m \sim 1.4\,M_\odot$, tidal stability in the gravitational field of Sgr~A* imposes an additional constraint on any bound dusty envelope. The characteristic size of the region within which material can remain gravitationally bound to the compact object is given by the Hill radius,
\begin{equation}
R_{\rm H} \simeq r \left(\frac{m}{3M_\bullet}\right)^{1/3},
\end{equation}
where $r$ is the orbital distance from Sgr~A* and $M_\bullet \simeq 4.3 \times 10^6\,M_\odot$ is the mass of the central supermassive black hole. Evaluating this expression at the observed pericenter distance of G2, $r_p \simeq 2000\,R_s \approx 170\,\mathrm{AU}$ \cite{Gillessen:2013G2}, yields
\begin{equation}
R_{\rm H}(r_p) \approx 0.8\,\mathrm{AU}
\end{equation}
for a $\sim$solar-mass object. The inferred blackbody radius of order $\sim 1\,\mathrm{AU}$ therefore lies slightly below, the tidal stability limit at pericenter. This is especially true for the grey body fit which resulted in a dust radius of 3.75AU and using the orbital parameters in \cite{Ciurlo:2020Nature} results in a Hill radius of $r_H \approx 18 $ AU which would be consistent with the infrared emission remaining compact during periapse. This further emphasizes that any viable model must account not only for the internal heating requirement but also for the strong dynamical constraints imposed by the Galactic Center environment.

\subsection{Dust Reprocessing of Accretion Luminosity}
To account for the H excess, \citet{Peissker2020}  proposed a two-component model consisting of a central object surrounded by a compact dusty envelope, with the envelope dominating the infrared emission. In the present work, we adopt the same physical requirement—namely that the dust must be heated internally—and investigate whether accretion onto a stellar-mass black hole can provide sufficient energy input to sustain the observed dust temperatures.
The bolometric accretion luminosity is determined by the radiative
efficiency $\epsilon$ and the mass accretion rate $\dot{M}_{\rm B}$,
\begin{equation}
L_{\rm bol} \equiv \epsilon\, \dot{M}_{\rm B} c^{2},
\end{equation}
where $\dot{M}_{\rm B}$ denotes the mass capture accretion rate onto the compact object.

To express the luminosity and accretion rate in dimensionless form,
we define
\begin{equation}
\ell \equiv \frac{L_{\rm bol}}{L_{\rm Edd}},
\qquad
\dot{m} \equiv \frac{\dot{M}_{\rm B}}{\dot{M}_{\rm Edd}},
\end{equation}
where
\begin{equation}
L_{\rm Edd} = 1.26 \times 10^{38}
\left( \frac{M_{\rm bh}}{M_\odot} \right)
\,{\rm erg\,s^{-1}}
\end{equation}
is the Eddington luminosity and the Eddington accretion rate is
\begin{equation}
\dot{M}_{\rm Edd} \equiv \frac{L_{\rm Edd}}{ c^{2}},
\end{equation}
These definitions imply $\ell = \epsilon \, \dot{m}$ and following \cite{Flores:2023PRD} we can parametrize $\epsilon = \eta \dot{m} $ such that $\eta$  encapsulated several accretion mechanisms as in \cite{ParkOstriker2001}. The dimensionless luminosity is then given by \cite{Murray_Clay_2012}:

\begin{equation}
\ell = \, \eta \,  \dot{m}^{2},
\end{equation}
Other ADAF solutions such as those in \cite{Beckert_2002} however show a somewhat steeper scaling with $\ell \propto \mdot ^{3.3} $.
We assume that the accretion luminosity is absorbed and reprocessed by dust located at a characteristic radius $R_d$ from the compact object. Imposing radiative equilibrium, the dimensionless luminosity required to sustain a dust temperature $T_d$ is

\begin{equation}
\ell
=
\frac{4\pi R_{\rm eff}^2 \sigma_{\rm SB} T_d^4}
{C_f \, L_{\rm Edd}}.
\end{equation}

where $\sigma_{\rm SB}$ is the Stefan--Boltzmann constant and $C_f$ 
denotes the dust covering fraction and $R_{\rm eff}$ is the effective dust radius defined as $R_{\rm eff}^2 \equiv C_f R_d^2$ where $R_{\rm eff}$ is the radius calculated from the fits.

For G2 ($T=623.23\,{\rm K}$, $R_{\rm eff}=0.5043\,{\rm AU}$),
\begin{equation}
\ell_{\rm G2}(C_f)=
\frac{4.86\times10^{-5}}{C_f}\left(\frac{M_\odot}{M_{\rm bh}}\right).
\end{equation}

For G3 ($T=515.35\,{\rm K}$, $R_{\rm eff}=1.2870\,{\rm AU}$),
\begin{equation}
\ell_{\rm G3}(C_f)=
\frac{1.48\times10^{-4}}{C_f}\left(\frac{M_\odot}{M_{\rm bh}}\right).
\end{equation}
Substituting the required dimensionless luminosities for G2 and G3 into the relation $\dot M_{\rm B} = (L_{\rm Edd}/c^2)\sqrt{\ell/\eta}$ yields the following accretion rates in terms of the dust covering fraction $C_f$ and the efficiency parameter $\eta$:

\begin{equation}
\dot M_{\rm B,G2}
\simeq 1.55\times10^{-11}\,
\left(C_f\,\eta\right)^{-1/2}
\;M_\odot\,{\rm yr^{-1}},
\end{equation}

\begin{equation}
\dot M_{\rm B,G3}
\simeq 2.70\times10^{-11}\,
\left(C_f\,\eta\right)^{-1/2}
\;M_\odot\,{\rm yr^{-1}}.
\end{equation}

\begin{figure*}[ht]
    \centering
    \includegraphics[width=1\textwidth]{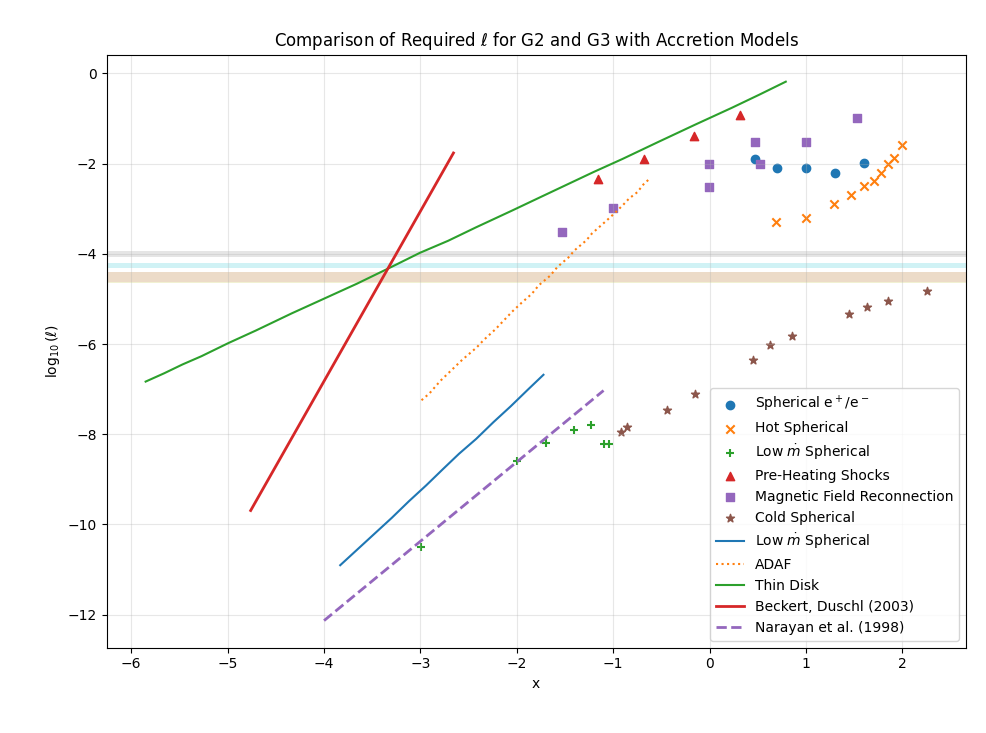}
\caption{
Dimensionless luminosity $\ell = L_{\rm bol}/L_{\rm Edd}$ as a function of accretion rate $\dot m$. 
Literature accretion solutions (lines and symbols) taken from \cite{ParkOstriker2001, Beckert_2002, narayan1998}  are compared with values derived from the G2 blackbody(brown band) and G3 greybody(grey band) and blackbody(blue band) fits. 
}
    \label{fig:Lumi}
\end{figure*}

In Fig. ~\ref{fig:Lumi}, we plot the implied dimensionless luminosities and compare them with black hole accretion solutions from \cite{ParkOstriker2001, Beckert_2002, narayan1998}. We find that the BH+Dust envelope model is consistent with these accretion solutions, with G2 and G3 producing very similar dimensionless luminosities. Notably, the models remain well matched to ADAF solutions even as the covering fraction varies, which would require higher $\ell $ values. This consistency motivates our subsequent analysis of the $\rm Br,\gamma$ emission in the following section.

\subsection{PBH + gas Halo}
To evaluate whether steady accretion from the ambient Galactic Center medium can provide sufficient heating, we compute the Bondi--Hoyle--Lyttleton (BHL) accretion rate along a G2-like orbit using observationally motivated radial profiles for both the gas density and temperature. We adopt the density profile \citep{Xu2006},
\begin{equation}
n(r) \simeq 10^{4}
\left(\frac{r}{10^{15}\,\mathrm{cm}}\right)^{-1}
\mathrm{cm^{-3}},
\end{equation}
and assume a hot Galactic Center atmosphere with temperature scaling\cite{Xu2006}
\begin{equation}
T(r) = 1.2 \times 10^8
\left(\frac{r}{1.4 \times 10^4 R_S}\right)^{-1},
\end{equation}
consistent with hydrostatic models of the inner accretion flow.

For an orbit with eccentricity $e=0.98$, the instantaneous velocity is obtained from the vis-viva relation,
\begin{equation}
v(r)=\sqrt{GM_\bullet\left(\frac{2}{r}-\frac{1}{a}\right)},
\end{equation}
where $a$ is determined from the orbital period via Kepler’s third law.

The accretion rate is computed as
\begin{equation}
\dot M_{\rm BHL}(r)=
4\pi G^2 M^2
\frac{\rho(r)}
{\left[v_{\rm rel}(r)^2+c_s(r)^2\right]^{3/2}},
\end{equation}
with $\rho(r)=\mu m_p n(r)$ and
\begin{equation}
c_s(r)=\sqrt{\frac{\gamma k_B T(r)}{\mu m_p}}.
\end{equation}
We take $v_{\rm rel}\simeq v(r)$, appropriate for a compact object moving through the ambient medium.

\begin{figure}[t]
\centering
\includegraphics[width=0.9\linewidth]{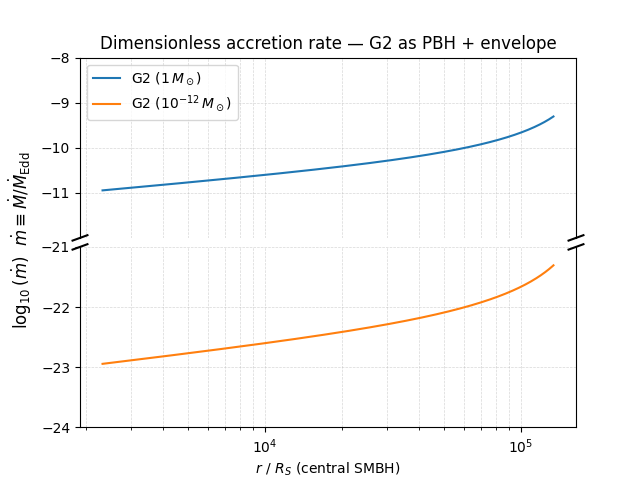}
\caption{
Bondi--Hoyle--Lyttleton accretion rate $\dot M(r)$ and corresponding dimensionless rate $\dot m=\dot M/\dot M_{\rm Edd}$ computed along a G2-like orbit with $e=0.98$ in the observed Galactic Center density profile. Throughout its orbit the accretion rate remains $\dot m\sim10^{-10}$--$10^{-11}$ for $1$--$2\,M_\odot$ black holes, implying extremely low radiative output in the RIAF regime.
}
\label{fig:mdot_profile}
\end{figure}

Figure~\ref{fig:mdot_profile} shows the resulting $\dot M(r)$ and corresponding dimensionless accretion rate $\dot m=\dot M/\dot M_{\rm Edd}$ for $1$--$2\,M_\odot$ black holes. Across the entire orbit, including at periapse, we obtain
\begin{equation}
\dot m \sim 10^{-10}\text{--}10^{-11}.
\end{equation}
In the radiatively inefficient regime appropriate for such extremely low accretion rates, the luminosity scales as $\ell=\eta \dot m^{2}$. 
For $\dot m \sim 10^{-10}$--$10^{-11}$, this implies
\begin{equation}
\ell \lesssim 10^{-20},
\end{equation}
even for $\eta \sim 1$, which is $\gtrsim 15$ orders of magnitude below the $\ell \sim 10^{-5}$--$10^{-4}$ required to sustain the observed dust temperatures $T\sim500\,\mathrm{K}$ inferred from the SED fits (Fig.~\ref{fig:Lumi}). 
Steady Bondi--Hoyle accretion from the ambient Galactic Center medium is therefore energetically incapable of powering the infrared emission.

Moreover, the corresponding mass supply is negligible. With $\dot M_{\rm BHL} \sim 10^{-20}\,M_\odot\,\mathrm{yr^{-1}}$, the total mass accumulated over a timescale $t \sim 10^{4}\,\mathrm{yr}$ would be only
\begin{equation}
M_{\rm acc} \sim \dot M_{\rm BHL} t \sim 10^{-16}\,M_\odot,
\end{equation}
many orders of magnitude below that required to assemble or maintain an AU-scale dusty envelope with appreciable optical depth. 
We therefore conclude that steady accretion from the ambient hot medium can neither generate sufficient internal heating nor build up a substantial dusty envelope around a stellar-mass compact object.

In contrast, neutron-star destruction naturally supplies a substantial reservoir of ejecta with mass $M_{\rm ej} \sim 10^{-4}$--$10^{-3}\,M_\odot$ (Fig.~\ref{fig:Mej_vs_radius}). If even a modest fraction of this material remains gravitationally bound, it can form a compact, optically thick envelope surrounding the compact remnant. In this case, the relevant mass supply for accretion is no longer the negligible Bondi--Hoyle rate from the ambient Galactic Center medium, but rather the bound ejecta itself. The accretion rate is therefore regulated by the envelope mass and its internal dynamics, not by the extremely low density of the surrounding hot atmosphere. This naturally explains how G-objects may remain infrared-bright despite the vanishingly small ambient steady-state 
accretion rate.

\subsection{Ionizing photon budget and Br$\gamma$ emission}

A defining observational characteristic of G objects is the presence of compact Br$\gamma$ emission, tracing ionized gas that is spatially and spectrally distinct from the dust continuum. In the solar-mass black hole interpretation, this emission arises naturally from photoionization of a bound gaseous envelope by radiation from the accretion flow. We therefore compute the ionizing photon output of the accreting compact object and relate it directly to the expected Br$\gamma$ luminosity.

\subsubsection*{Ionization--bounded regime}

In the ionization--bounded (photon--limited) regime, the gas reservoir is sufficiently large that essentially all hydrogen-ionizing photons are absorbed within the nebula. Photoionization equilibrium then requires that the total recombination rate balance the ionizing photon production rate,
\begin{equation}
Q_H = \int \alpha_B(T)\, n_e n_p \, {\rm d}V ,
\end{equation}
where $Q_H$ is the rate of photons with $h\nu > 13.6\,{\rm eV}$ and $\alpha_B(T)$ is the Case--B recombination coefficient.

Following \cite{Schartmann2015, Shaqil_2025}, we write the Case--B Br$\gamma$ volume emissivity as
\begin{equation}
j_{{\rm Br}\gamma}(T) = \epsilon_{{\rm Br}\gamma}(T)\, n_e n_p ,
\end{equation}
with
\begin{equation}
\epsilon_{{\rm Br}\gamma}(T)
=
3.44\times10^{-27}
\left(\frac{T}{10^4\,{\rm K}}\right)^{-1.09}
\;{\rm erg\ s^{-1}\ cm^{3}} .
\end{equation}
The total Br$\gamma$ luminosity is therefore
\begin{equation}
L_{{\rm Br}\gamma}
=
\int j_{{\rm Br}\gamma}\,{\rm d}V
=
\epsilon_{{\rm Br}\gamma}(T)
\int n_e n_p\,{\rm d}V .
\end{equation}
Eliminating the emission measure using the equilibrium condition yields
\begin{equation}
L_{{\rm Br}\gamma}
=
\frac{\epsilon_{{\rm Br}\gamma}(T)}{\alpha_B(T)}\,
Q_H .
\end{equation}
In this regime the Br$\gamma$ luminosity is set entirely by the ionizing photon budget and is independent of the detailed gas distribution.

\subsection{Ionizing photon rate from ADAF SEDs}

To compute $Q_H$ we adopt radiatively inefficient accretion flow (ADAF) spectral energy distributions (SEDs) from \citet{narayan1998} as well as \citet{Beckert_2002} which was based on the self-similar ADAF solutions of \citet{Narayan_1994,Beckert2000}. ADAFs are geometrically thick, optically thin, quasi-spherical accretion flows \cite{Narayan_1995} in which a large fraction of the viscously dissipated energy is advected inward with the gas and ultimately deposited into the black hole rather than radiated away. As a result, their radiative efficiencies are substantially lower than those of standard thin disks.

The emergent ADAF spectra are composed of three primary radiative components. At radio frequencies the emission is dominated by thermal synchrotron radiation from hot electrons, producing a characteristic power-law spectrum with spectral index $\alpha = 2/5$ ($F_\nu \propto \nu^\alpha$). At higher frequencies (mm to X-ray), the emission is governed by inverse Compton scattering of synchrotron seed photons and thermal bremsstrahlung, with the relative importance of these components determined by parameters such as $\dot m$,the electron temperature and optical depth \cite{Mahadevan1997}. For each SED we compute its bolometric luminosity directly from the spectral shape via

\begin{equation}
L_{\rm bol,ref}(\dot m)
=
\int_0^\infty \nu L_\nu \, {\rm d}\ln\nu ,
\end{equation}

and the corresponding hydrogen-ionizing photon production rate

\begin{equation}
Q_{H,{\rm ref}}(\dot m)
=
\int_{\nu>\nu_H}
\frac{\nu L_\nu}{h\nu}
\,{\rm d}\ln\nu ,
\end{equation}

where $h\nu_H = 13.6\,{\rm eV}$. To apply the same SED family to a different black hole mass $M$, we adopt the ADAF normalization scaling $\nu L_\nu \propto M$ at fixed $\dot m$\cite{Beckert_2002}.  Under this assumption, both the bolometric luminosity and the ionizing photon rate scale linearly with mass:

\begin{equation}
\begin{split}
L_{\rm bol}(\dot m,M)
=
L_{\rm bol,ref}(\dot m)
\left(
\frac{M}{M_{\rm ref}}
\right),
\\
Q_H(\dot m,M)
=
Q_{H,{\rm ref}}(\dot m)
\left(
\frac{M}{M_{\rm ref}}
\right),  
\end{split}
\end{equation}

where $M_{\rm ref}$ is the black hole mass corresponding to the tabulated SEDs. 
This procedure rescales only the luminosity normalization; the spectral shape in frequency space is left unchanged.

The predicted Br$\gamma$ luminosity in the ionization-bounded regime is then

\begin{equation}
L_{{\rm Br}\gamma}
=
\frac{\epsilon_{{\rm Br}\gamma}(T)}{\alpha_B(T)}
\,Q_H(\dot m,M).
\end{equation}

The observed integrated line flux is
\begin{equation}
F_{\rm line}
=
\frac{L_{{\rm Br}\gamma}}{4\pi D^2},
\end{equation}
and assuming a Gaussian line profile with FWHM $\Delta v$, the peak flux density is
\begin{equation}
F_{\nu,{\rm peak}}
=
\frac{F_{\rm line}}{1.064\,\Delta\nu},
\qquad
\Delta\nu
=
\nu_{{\rm Br}\gamma}
\left(
\frac{\Delta v}{c}
\right).
\end{equation}

\begin{figure}[t]
\centering
\includegraphics[width=1\linewidth]{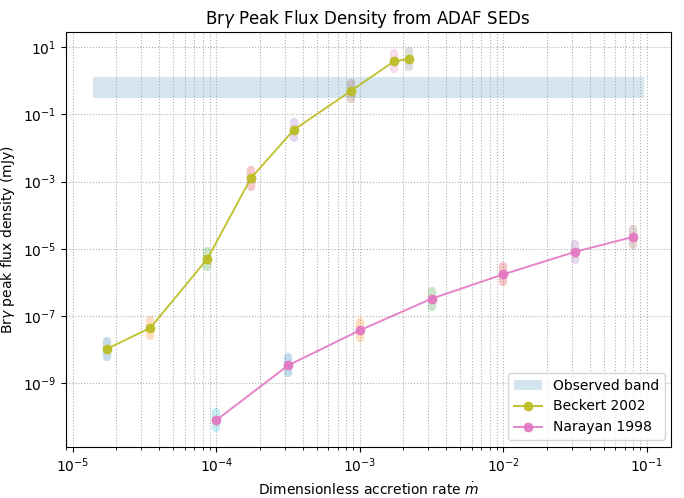}
\caption{ Bry$\gamma$ luminosity as a function od the dimensionless accretion rate, $\dot m $. The SEDs used are those of the ADAF solutions found in \citet{Beckert_2002} and \citet{narayan1998}. The maximum $\mdot$ is that which corresponds to $\dot m_{\rm crit}$ above which there is no longer an ADAF solution.
}
\label{fig:Bry_ADAF}
\end{figure}

Using the observed Br$\gamma$ peak flux densities \cite{Ciurlo:2020Nature} and adopting $\Delta v\in[100,300]\,{\rm km\,s^{-1}}$. The necessay $\dot m $ to satisfy the Br$\gamma$ observations is shown in Fig ~(\ref{fig:Bry_ADAF}) demonstrating that an ADAF with $ \dot m \gtrsim 10^{-3}$ can sustain the ionization and furthermore we infer that viable accretion models must produce ionizing photon rates of $Q_H \sim 10^{42} - 10^{44}\ {\rm s^{-1}}.$

Our calculation implicitly assumes that all ionizing photons emitted by the accretion flow are available to photoionize the gaseous envelope. In practice, ionizing radiation must propagate through a dusty, optically thick reprocessing layer whose EUV opacity can strongly suppress the delivered Lyman-continuum photon rate. We therefore distinguish between the intrinsic ionizing photon production rate, $Q_H^{\rm int}$, and the effective rate that reaches the gas, $Q_H^{\rm eff} \equiv f_{\rm esc,H} Q_H^{\rm int}$, where $f_{\rm esc,H}$ encapsulates absorption by dust and/or escape from the system. In the ionization-bounded regime, the predicted recombination-line luminosity becomes
\begin{equation}
L_{{\rm Br}\gamma}
=
\frac{\epsilon_{{\rm Br}\gamma}(T)}{\alpha_B(T)}\, Q_H^{\rm eff}
=
\frac{\epsilon_{{\rm Br}\gamma}(T)}{\alpha_B(T)}\, f_{\rm esc,H}\, Q_H^{\rm int}.
\end{equation}
For a uniform dusty screen one expects $f_{\rm esc,H}\sim \langle e^{-\tau_\nu}\rangle_{\nu>\nu_H}$, which can be exponentially small for modest EUV optical depths. However, if the envelope is clumpy and porous, ionizing photons may escape through low-column channels. In the limit where dusty clumps are optically thick to the EUV, the delivered ionizing fraction is set primarily by the open solid-angle fraction, $f_{\rm esc,H}\approx f_\Omega$, allowing a non-negligible $Q_H^{\rm eff}$ even when most of the luminosity is reprocessed into the infrared. Lastly we observe that the slopes in the $L_{\rm Br \gamma} - \dot m$ and $\ell - \dot m $ are very similar which would lead one leads us to conclude that a number of different ADAFs can produce the necessary $L_{\rm Br \gamma}$ with the main limiting factor being the optical depth and coverage of the dusty envelope which would likely a dominant fraction absent fine tuning.

\subsection{X-ray transmission through a compact envelope.}
For $\mdot \lesssim 10^{-3}$ the the X-ray emission is dominated by bremsstrahlung such that $L_X \propto m \dot m^2$ \cite{Yi1998ApJ,Mahadevan_1997} whereas when $\dot m > 10^{-3}$ the luminosity is dominated by Compton scattering \cite{Yi1998ApJ,Mahadevan_1997}
The intrinsic \(2\!-\!8~{\rm keV}\) luminosity follows from
\begin{equation}
L^{\rm int}_{2-8}=\int_{2~{\rm keV}}^{8~{\rm keV}} \nu L_\nu\,{\rm d}\ln\nu.
\end{equation}
To model attenuation by a surrounding gas column \(N_{\rm H}\), we apply photoelectric absorption to the spectral energy distribution \cite{Wilms2000},
\begin{equation}
L^{\rm trans}_{2-8}(N_{\rm H})=
\int_{2~{\rm keV}}^{8~{\rm keV}} \nu L_\nu\,
\exp\!\left[-N_{\rm H}\,\sigma_{\rm pe}(E)\right]\,{\rm d}\ln\nu,
\end{equation}
with \(\sigma_{\rm pe}(E)\) the photoelectric cross-section per H nucleus (we adopt a standard neutral-gas approximation \(\sigma_{\rm pe}\propto E^{-3}\)\cite{Morrison1983} in this band for an order-of-magnitude transmission estimate). We then plot \(\log_{10}L^{\rm trans}_{2-8}\) as a function of \(N_{\rm H}\) for each \(\dot m\), providing a direct mapping between intrinsic accretion-powered X-ray output and observable transmitted luminosity.

\paragraph{Column-density scaling from an envelope mass and radius.}
For an approximately uniform envelope of gas mass $M_{\rm gas}$ and characteristic radius $R$, the characteristic hydrogen column density is
\begin{equation}
N_{\rm H}
\simeq
1.2\times10^{24}
\left(\frac{M_{\rm gas}}{10^{-4}M_\odot}\right)
\left(\frac{R}{10~{\rm AU}}\right)^{-2}
~{\rm cm^{-2}} .
\end{equation}
Such columns strongly attenuate hard X-rays in the $2\!-\!8\,\mathrm{keV}$ band and therefore provide a direct connection between envelope compactness and observable X-ray output.

Deep \textit{Chandra} observations of the inner Galactic Center reach a point-source sensitivity of order 
$L_{2-8} \lesssim 10^{30}\,\mathrm{erg\,s^{-1}}$ at 8\,kpc in the ultra-deep $4.5$\,Ms survey of the inner $500''$ 
\citep{Zhu_2018,Muno_2003}. No X-ray counterpart has been identified at the position of G2 during extensive 
monitoring of the Galactic Center \citep[e.g.][]{Haggard2014,Bouffard2019}, implying that any compact object 
embedded within the dusty envelope must satisfy
\begin{equation}
L^{\rm trans}_{2-8} \lesssim 10^{31}\,\mathrm{erg\,s^{-1}} .
\end{equation}

\begin{figure}[t]
\centering
\includegraphics[width=1\linewidth]{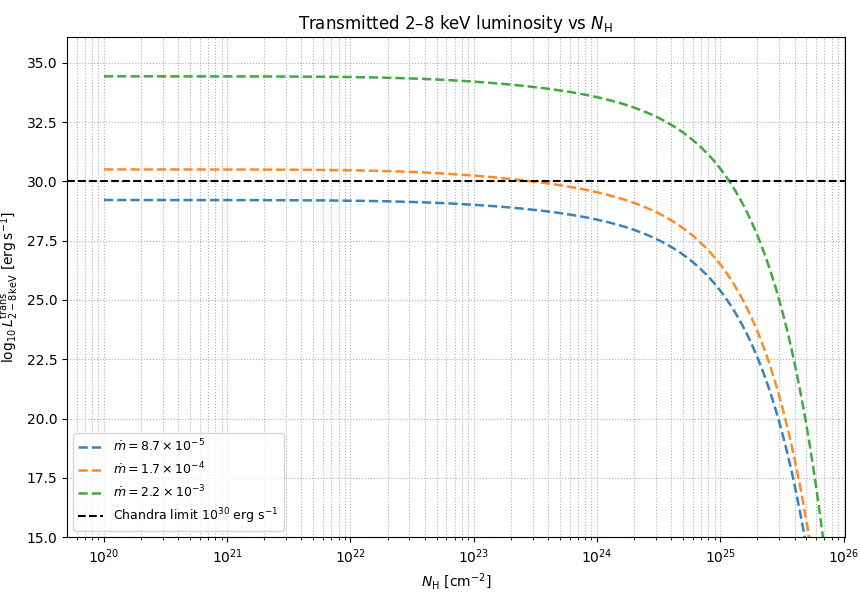}
\caption{
Transmitted $2\!-\!8\,\mathrm{keV}$ luminosity as a function of envelope column density $N_{\rm H}$ for representative accretion rates using the SED from \citep{Beckert_2002}. The horizontal line marks the approximate \textit{Chandra} point-source sensitivity in the Galactic Center ($L_{2-8} \sim 10^{30}\,\mathrm{erg\,s^{-1}}$; \citealt{Muno_2003,Zhu_2018}). Only models with sufficiently large $N_{\rm H}$ (compact, high-mass envelopes) fall below the observational upper limit.
}
\label{fig:X-ray}
\end{figure}

In radiatively inefficient accretion flows, the hard X-ray luminosity scales approximately as  $L_X \propto M \dot m^2$ \citep{Mahadevan1997,Yi1998ApJ}. For a $\sim 1\text{--}2\,M_\odot$ black hole, accretion rates $\dot m \gtrsim 10^{-3}$ typically produce intrinsic luminosities  $L^{\rm int}_{2-8} \sim 10^{33}\text{--}10^{34}\,\mathrm{erg\,s^{-1}}$, well above the \textit{Chandra}  non-detection threshold. As shown in Fig.~\ref{fig:X-ray}, suppressing the transmitted luminosity below the observational limit requires columns  $N_{\rm H} \gtrsim 10^{24}\text{--}10^{25}\,\mathrm{cm^{-2}}$, corresponding to $M_{\rm gas} \sim 10^{-4}\,M_\odot$ confined within radii $\lesssim 10\,\mathrm{AU}$ for representative parameters.  The X-ray non-detection therefore imposes a stringent joint constraint: either the accretion rate must be  $\dot m \ll 10^{-3}$, or the envelope must be sufficiently compact and massive to remain optically thick in the hard X-ray band.

\subsection{Radio Continuum and Hydrogen Recombination Lines}
\label{sec:radio}

The ionized envelopes surrounding PBH--NS remnants are expected to emit thermal free--free radiation in the radio band. These ``skins'' are maintained by photoionization from the accretion luminosity and collisional ionization from hot electrons in the surrounding medium \cite{Draine2011}. The resulting emission is characterized by nearly flat spectra and weak hydrogen recombination lines, providing a complementary diagnostic to the NIR features discussed in Sec.~\ref{sec:observational_status}.

\paragraph{Free--free emission.}
For an approximately homogeneous ionized sphere of radius $R_{\rm env}$, electron temperature $T_e$, and density $n_e$, the specific luminosity is
\begin{equation}
L_\nu^{\rm ff} \simeq 4\pi\,6.8\times10^{-38}\,Z^2\,n_e^2\,T_e^{-1/2}\,
   e^{-h\nu/kT_e}\,\bar g_{\rm ff}\,V,
\end{equation}
where $\bar g_{\rm ff}\!\sim\!1.1\!-\!1.5$ is the Gaunt factor and $V=(4\pi/3)R_{\rm env}^3$.
The optical depth at frequency $\nu$ is
\begin{equation}
\tau_\nu \simeq 0.082\,T_e^{-1.35}\,
\left(\frac{\nu}{{\rm GHz}}\right)^{-2.1}
\left(\frac{{\rm EM}}{{\rm pc\,cm^{-6}}}\right),
\end{equation}
with emission measure ${\rm EM}=n_e^2R_{\rm env}$.  
At GHz frequencies and $T_e\simeq10^4$~K, the envelopes are generally marginally optically thin ($\tau_\nu\lesssim1$), producing brightness temperatures $T_b \approx T_e(1-e^{-\tau_\nu})$ and flux densities
\begin{equation}
\begin{split}
\label{flux_eq}
S_\nu \approx
3.1\,{\rm mJy}&
\left(\frac{T_e}{10^4\,{\rm K}}\right)^{-0.35}
\left(\frac{\nu}{10\,{\rm GHz}}\right)^{-0.1}\\& \times
\left(\frac{{\rm EM}}{10^{8}\,{\rm pc\,cm^{-6}}}\right)
\left(\frac{\theta_{\rm s}}{0.1''}\right)^{2},
\end{split}
\end{equation}
where $\theta_{\rm s}$ is the angular size at 8~kpc.
The predicted spectral index $\alpha\simeq-0.1$ to $-0.3$ ($S_\nu\!\propto\!\nu^\alpha$) is consistent with flat radio sources observed toward the Galactic Center \citep{Mori2013}.

\paragraph{Hydrogen radio recombination lines.}
Hydrogen RRLs such as H30$\alpha$ (231~GHz), H41$\alpha$ (92~GHz), and H92$\alpha$ (8.3~GHz) may accompany the continuum, with line-to-continuum ratios
\begin{equation}
\frac{\Delta S_{\nu,{\rm line}}}{S_{\nu,{\rm cont}}}\approx
2.0\times10^3\,T_e^{-1.15}\,
\left(\frac{\Delta v}{{\rm km\,s^{-1}}}\right)^{-1},
\end{equation}
typically $10^{-3}$--$10^{-2}$ for $T_e\simeq10^4$~K and line widths $\Delta v\simeq100$--300~km~s$^{-1}$ \cite{Ciurlo:2020Nature}.
Detection of these lines with ALMA, VLA, or MeerKAT \cite{Royster_2019,Lang_2001, Emig_2023} would provide direct evidence for ionized gas and enable velocity measurements tracing orbital motion.

\paragraph{Environmental dependence.}
The radio luminosity depends sensitively on local conditions:
\begin{itemize}
\item \textbf{Density structure.}
For envelope densities $n_e \sim 10^5$--$10^6~{\rm cm^{-3}}$ and characteristic radii of tens of AU, the resulting emission measures ${\rm EM} \sim 10^7$--$10^9~{\rm pc\,cm^{-6}}$ produce sub-mJy flux densities at $\nu \sim 10$~GHz for sources at 8~kpc, placing them within reach of deep VLA observations.  While Eq.~\ref{flux_eq} assumes a uniform sphere, realistic envelopes are likely stratified, with density profiles $n_e(r)\propto r^{-\beta}$.  Because the free--free emissivity scales as $j_\nu^{\rm ff}\propto n_e^2$, centrally concentrated halos can yield significantly larger emission measures than uniform configurations with the same total gas mass \cite{Osterbrock2006,Draine2011}.  Compact, centrally peaked envelopes are therefore strongly favored for producing detectable radio flux.
\item \textbf{Temperature.}
In the optically thin regime, $S_\nu \propto T_e^{-0.35}$, reflecting the weak temperature dependence of free--free emission.  Electron temperatures in the ionized streamers of Sgr~A West (the mini-spiral) are typically $T_e \sim (5$--$13)\times10^3$~K, i.e. of order $10^4$~K \citep{Zhao_2009}.  Thus, even factor-of-two variations in $T_e$ change the predicted radio flux by only $\sim20$--30\%, substantially less than uncertainties in density structure or envelope size. 
\item \textbf{Velocity field:} 
Motion of the source through the ambient Galactic Center medium can, in principle, generate ram-pressure compression, forming a bow shock at the leading edge or a cometary tail in the downstream wake. Such structures enhance the local emission measure ($\propto n_e^2$) and were predicted to produce detectable radio emission during G2’s pericentre passage, either through direct interaction with the ambient medium or via a wind-driven shock (e.g., \citealt{Narayan_2012,CrumleyKumar2013}).  However, no significant radio brightening was observed during or after pericentre \citep{Bower_2015}, and models have likewise struggled to reproduce the relatively flat Br$\gamma$ light curve without invoking fine tuning.  In particular, the absence of a detectable radio flare implies that any ionized envelope must be sufficiently compact a condition naturally satisfied in the PBH--NS remnant scenario.
\end{itemize}

\paragraph{Observational strategy.}
\begin{enumerate}
\item \emph{Continuum imaging:}  
Target known G objects with VLA S/X bands or MeerKAT L band.  The angular broadening inferred from the Galactic Center magnetar SGR~J1745$-$2900 allows compact sources ($\lesssim0.1''$ at 8~GHz) to remain detectable through scattering \citep{Mori2013}.
\item \emph{Spectral index tests:}  
Multi-frequency coverage (5--15~GHz) can test for the predicted $\alpha\!\approx\!-0.1$--$-0.3$, discriminating free--free emission from stellar winds ($\alpha\simeq0.6$) \cite{WrightBarlow1975} or synchrotron sources.
\item \emph{Recombination-line spectroscopy:}  
High-resolution ($\Delta v\!\lesssim\!50$~km~s$^{-1}$) ALMA follow-up in H30$\alpha$ or H41$\alpha$ can constrain $n_e$, $T_e$, and kinematics.
\end{enumerate}

\paragraph{Interpretation.}
Flat-spectrum ($\alpha\!\approx\!-0.2$), sub-mJy sources coincident with IR-identified G objects would provide compelling evidence for thermally ionized envelopes surrounding low-efficiency accreting compact remnants.
Their radio flux levels, spectral slopes, and lack of non-thermal counterparts would distinguish them from T~Tauri stars or massive YSOs\cite{Scoville2013}, supporting the PBH--NS conversion interpretation.

\begin{figure}[t]
\centering
\includegraphics[width=1\columnwidth]{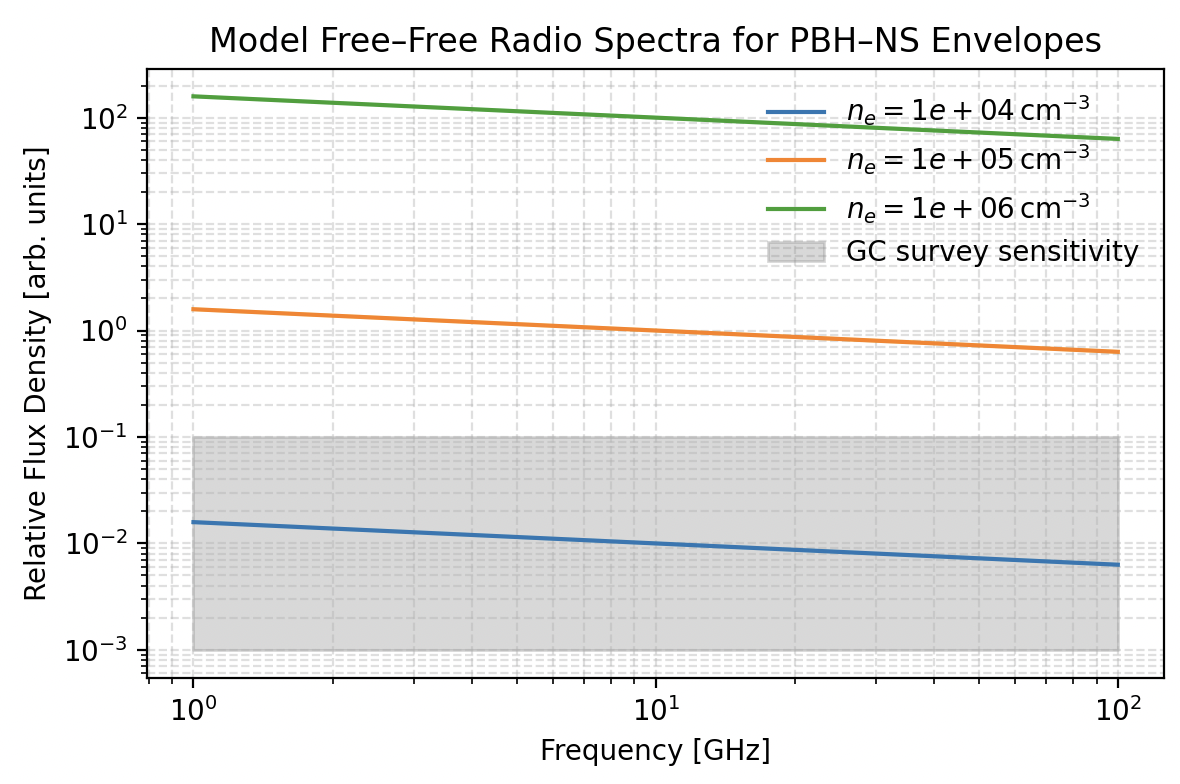}
\caption{Model radio spectra from thermal free--free emission for ionized envelopes with $T_e=10^4$~K and $R_{\rm env}=50$~AU. 
Lines correspond to electron densities $n_e = 10^4$, $10^5$, and $10^6$~cm$^{-3}$.
The shaded band shows approximate GC radio survey sensitivity limits.
The nearly flat slope ($S_\nu\propto\nu^{-0.2}$) is a distinctive signature of optically thin free--free emission from compact ionized envelopes.}
\label{fig:radio_spectrum}
\end{figure}

\subsection{Microlensing framework and event-rate calculation}

Gravitational microlensing provides a powerful probe of non-luminous and compact objects at astronomical distances and has been widely applied to the study of exoplanets and free-floating planets \cite{mao1991, gaudi2012, sumi2011unbound}, the measurement of stellar remnant masses \cite{gould1994, wyrzykowski2016}, and even exotic objects such as wormholes \cite{Gao_2023, Gao_2024, Liu_2023}. Microlensing is a purely gravitational phenomenon and is therefore sensitive to the mass of an object irrespective of its electromagnetic emission, making it particularly well suited for searches for macroscopic dark matter, and specifically PBHs, in the Milky Way \cite{paczynski1986}. Previous surveys such as OGLE \cite{Niikura_2019}, Subaru Hyper Suprime-Cam (HSC) \cite{niikura2019microlensing}, and the EROS/MACHO collaborations \cite{Alcock_1998} have largely ruled out brown dwarfs in the mass range \( \sim 10^{-7} - 10\,\Msun \), and have placed some of the strongest constraints on the mass and abundance of primordial black holes (PBHs) over the broad range \( \sim 10^{-10} - 10\,\Msun \). These results imply that PBHs cannot constitute the dominant component of dark matter across most of this mass range, leaving only a narrow window around \( \sim 10^{-16} - 10^{-11}\,\Msun \) in which PBHs could account for all of the dark matter. Notably, an excess of short-timescale microlensing events reported in \cite{Niikura_2019} has been proposed to arise from a population of Earth-mass PBHs (\( \sim 10^{-6}\,\Msun \)) \cite{Li_2025}. 

When a massive object (the lens) passes close to the line of sight to a background star (the source), its gravitational field bends the light rays, producing a transient magnification of the source. For sufficiently massive lenses, multiple images may form, while for low-mass lenses the images remain unresolved and appear as a single magnified source. The characteristic angular scale of a microlensing event is the Einstein angle,

\begin{equation}
\theta_E = \sqrt{\frac{4GM}{c^2}\left(1-\frac{d_L}{d_S}\right)\frac{1}{d_L}},
\end{equation}
which depends on the lens mass and the distances to the lens, $d_L$ and source, $d_S$ \cite{DeRocco_2024, DeRocco_2024b}. In the point-source regime (\(\theta_S \ll \theta_E\)), the event duration is set by the Einstein crossing time, while in the finite-source regime (\(\theta_S \gtrsim \theta_E\)) the peak magnification is suppressed and the duration is governed by the time required for the lens to traverse the source disk \cite{DeRocco_2024, DeRocco_2024b}. For most microlensing events—particularly those involving low-mass lenses or short timescales—the only robustly measurable observable is the event duration, \(t_{\rm dur}\), defined as the time over which the magnification exceeds a detection threshold.

\begin{figure}[t]
\centering
\includegraphics[width=1\columnwidth]{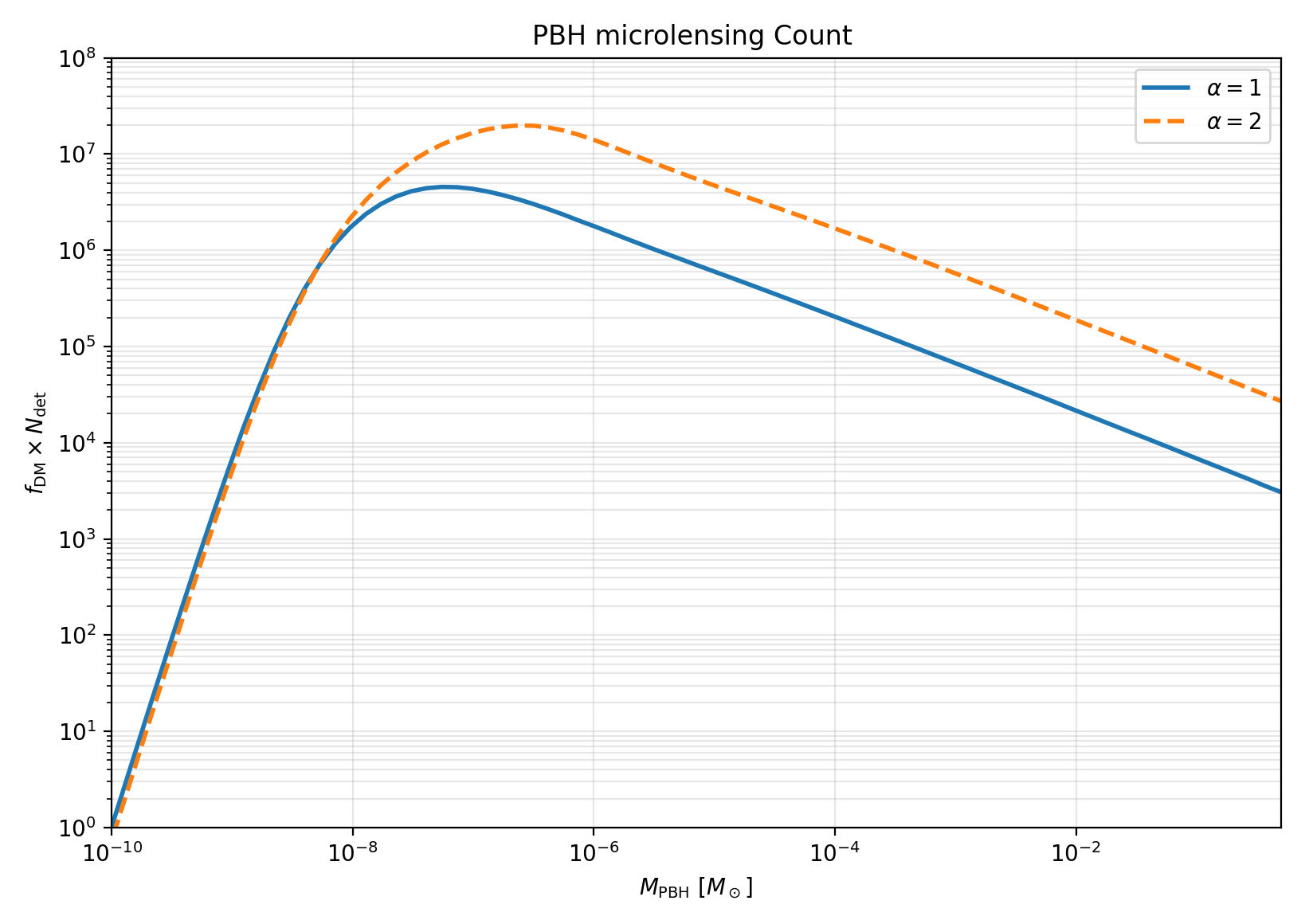}
\caption{Expected number of PBH microlensing detections for the Roman Space Telescope as a function of PBH mass \(m_{\rm PBH}\) for inner dark-matter slopes \(\alpha=1,2\). The steep drop at low masses reflects the finite-timescale detection threshold, corresponding to a minimum detectable mass \(M_{\rm cut}\).}
\label{fig:PBH_lensing_Roman}
\end{figure}

The number of microlensing events is computed by integrating the differential microlensing event rate over lens mass, distance, impact parameter, and duration. The differential event rate is given by \cite{Niikura_2019, DeRocco_2024}
\begin{equation}
\begin{split}
\frac{d\Gamma}{dM\,dd_L\,dt_{\rm dur}\,du_{\rm min}}
= &
\frac{2\,v_T^4}{u_T^2-u_{\rm min}^2}\,
\frac{1}{v_c^2}\, \\& \times
\exp\!\left(-\frac{v_T^2}{v_c^2}\right)
\frac{\rho_M}{M}\,f(M)\,\varepsilon(t_{\rm dur}),
\end{split}
\end{equation}
where \(f(M)\) is the lens mass function, \(\rho_M\) is the mass density of the lens population, and \(\varepsilon(t_{\rm dur})\) is the detection efficiency. The relative source-lens velocity $v_c$ is given by Eq \ref{v_rms} and transverse velocity, $v_T$ is related to the event duration via
\begin{equation}
v_T = \frac{2\,\theta_E d_L}{t_{\rm dur}}\sqrt{u_T^2-u_{\rm min}^2}.
\end{equation}
The total event rate is obtained by integrating this expression over the allowed parameter space, with the maximum impact parameter \(u_T\). The event rate is then:

\begin{align}
\label{microlensing_events}
\Gamma
&=
2
\int_{M_{\min}}^{M_{\max}} dM
\int_{0}^{d_S} d d_L
\int_{0}^{u_T} d u_{\min}
\int_{t_{\min}}^{t_{\max}} d t_{\rm dur}
\nonumber \\
&\qquad\times
\frac{1}{M}
\frac{1}{\sqrt{u_T^2 - u_{\min}^2}}
\frac{v_T^4}{v_c^2}
\exp\!\left(-\frac{v_T^2}{v_c^2}\right)
\rho_M
f(M)
\epsilon(t_{\rm dur}) \, ,
\end{align}
where for the monochromatic PBH mass distribution we consider here $f(M)$ is simply a delta function. With the event rate we can then calculate the number of microlensing events as:
\begin{equation}
    N_{\text{events}} = \Gamma \, \times n_{\text{sources}} \,  \times T_{\text{obs}}
\end{equation}
The upcoming Roman Space Telescope Galactic Bulge Time-Domain Survey will observe seven contiguous fields covering 2 square degrees of the Galactic bulge, yielding approximately $n_{\text{sources}} = 2.4 \times 10^8$ sources with a 15-minute cadence over six 72-day observing seasons \cite{Penny_2019, wilson2023transitingexoplanetyieldsroman}. Following \cite{DeRocco_2024} we define the threshold magnification as $A_T$ = 1.05. To solve the integral of Eq.~\eqref{microlensing_events} we use \textbf{LensCalc.py} a python package to calculate microlensing observables which is appropriate for population level studies \cite{DeRocco_2024}.
In Fig.~\ref{fig:PBH_lensing_Roman} we plot the expected number of PBH microlensing events as a function of the PBH mass \(m_{\rm PBH}\) for inner dark-matter slopes \(\alpha=1,2\), explicitly accounting for the finite-timescale detection threshold of Roman, which induces a low-mass cutoff which was summarized in as \citet{DeRocco_2024}: 
\begin{equation}
M_{\rm cut} \simeq \frac{(A_T^2-1)\,D_L R_\star^2 c^2}{16 D_S^2 G},
\end{equation}
where we have assumed \(D_L\ll D_S\); for representative values \(A_T=1.05\), \(D_L=1\,\mathrm{kpc}\), and \(D_S=8.5\,\mathrm{kpc}\), this corresponds to \(M_{\rm cut}\simeq 8.7\times10^{-10}\,M_\odot\), producing the sharp suppression of the event rate at low masses visible in the figure.

Our Bayesian analysis predicts a joint posterior over $\{M_{\rm PBH}, f_{\rm DM}, \alpha\}$, so the microlensing forecasts must be interpreted within this correlated parameter space rather than in terms of $M_{\rm PBH}$ alone. In standard photometric microlensing, the event timescale $t_E$ reflects a combination of lens mass, lens distance, and transverse velocity, so a single light curve typically constrains only this composite scale rather than uniquely determining the lens mass. Although Roman will be capable of astrometric microlensing measurements in principle, recent studies show that for low PBH masses ($M_{\rm PBH}\lesssim10^{-1}M_\odot$) the GBTDS cadence and precision are insufficient to robustly constrain microlensing observables for pure astrometric events \cite{Fardeen2023}. Since our posterior mass range extends only up to $\sim10^{-8}M_\odot$, astrometric mass determination is unlikely to substantially reduce degeneracies in the relevant regime. Within the regime relevant to our analysis, Roman microlensing should therefore be viewed primarily as a population-level probe whose constraining power arises from event statistics rather than per-event mass measurements. Because our neutron-star capture framework already restricts $\alpha$, $f_{\rm DM}$, and $M_{\rm PBH}$ within a correlated posterior, a statistically significant detection would reweight this parameter space and help localize the viable region of $\{M_{\rm PBH}, f_{\rm DM}\}$ consistent with both capture and microlensing observables. Such an interpretation must, however, be treated with caution, as short-timescale microlensing events may suffer from source confusion, particularly from free-floating planets(ffp), which can mimic low-mass compact-object signals and complicate a direct identification of PBHs\cite{DeRocco_2024b}. A more complete analysis would include the ffp background and perform a hierarchical mixture between the two populations. 

The broad posterior mass distribution plays a central role in shaping these expectations. A substantial fraction of posterior probability resides in the lower-mass mass region, below the Roman sensitivity threshold $M_{\rm cut}$, where the predicted number of detectable events is essentially zero; consequently, null detections are statistically favored. However, a small  high mass remains which traces the outer edge of the microlensing constraints is partially above $M_{\rm cut}$ and therefore contributes a finite probability of detectable events. We propagate the Galactic Center posterior
$p(\boldsymbol{\theta},\boldsymbol{\phi}\mid\mathcal D)$
into a Roman microlensing prediction through the posterior predictive distribution
\begin{equation}
p(N_{\rm det}\mid\mathcal D)
=
\int
\mathrm{Pois}\!\left(N_{\rm det}\mid\lambda(\boldsymbol{\theta})\right)
\,p(\boldsymbol{\theta}\mid\mathcal D)\,
d\boldsymbol{\theta},
\end{equation}
where $\lambda(\boldsymbol{\theta}) \equiv N_{\rm det}(\boldsymbol{\theta})$ denotes the predicted Roman event yield for
$\boldsymbol{\theta}=\{\alpha_{\rm DM},\gamma,M_{\rm PBH},f_{\rm DM}\}$.

Marginalizing over the GC posterior, we obtain
\begin{equation}
P(N_{\rm det} > 1 \mid \mathcal D) = 0.094
\end{equation}
Thus, only approximately $9.4\%$ of the GC-allowed parameter space predicts more than one detectable Roman microlensing event. To characterize the region of parameter space capable of generating multiple detections, we examine the subset of posterior draws satisfying $N_{\rm det} > 1$. 

Within this region we obtain the following marginalized constraints:
\begin{align*}
\alpha_{\rm DM}
&=
1.87
^{+0.10}_{-0.17},
\\[6pt]
\log_{10}(f_{\rm DM})
&=
-2.08
^{+0.60}_{-0.76},
\\[6pt]
\log_{10}\!\left(\frac{M_{\rm PBH}}{M_\odot}\right)
&=
-8.73
^{+0.81}_{-0.62}
\end{align*}
where uncertainties denote central $68\%$ credible intervals.

These values characterize the portion of the Galactic Center posterior that yields an observable Roman signal.If Roman observes an event count $N_{\rm obs}$, the Galactic Center posterior updates according to
\begin{equation}
p(\boldsymbol{\theta}\mid \mathcal D, N_{\rm obs})
\propto
\mathrm{Pois}\!\left(N_{\rm obs}\mid\lambda(\boldsymbol{\theta})\right)
\,p(\boldsymbol{\theta}\mid\mathcal D).
\end{equation}

In practice, this update is implemented by importance reweighting posterior samples drawn from $p(\boldsymbol{\theta}\mid\mathcal D)$ with weights
\begin{equation}
w(\boldsymbol{\theta})
\propto
\mathrm{Pois}\!\left(N_{\rm obs}\mid\lambda(\boldsymbol{\theta})\right).
\end{equation}

\begin{figure}[t]
\centering
\includegraphics[width=1\columnwidth]{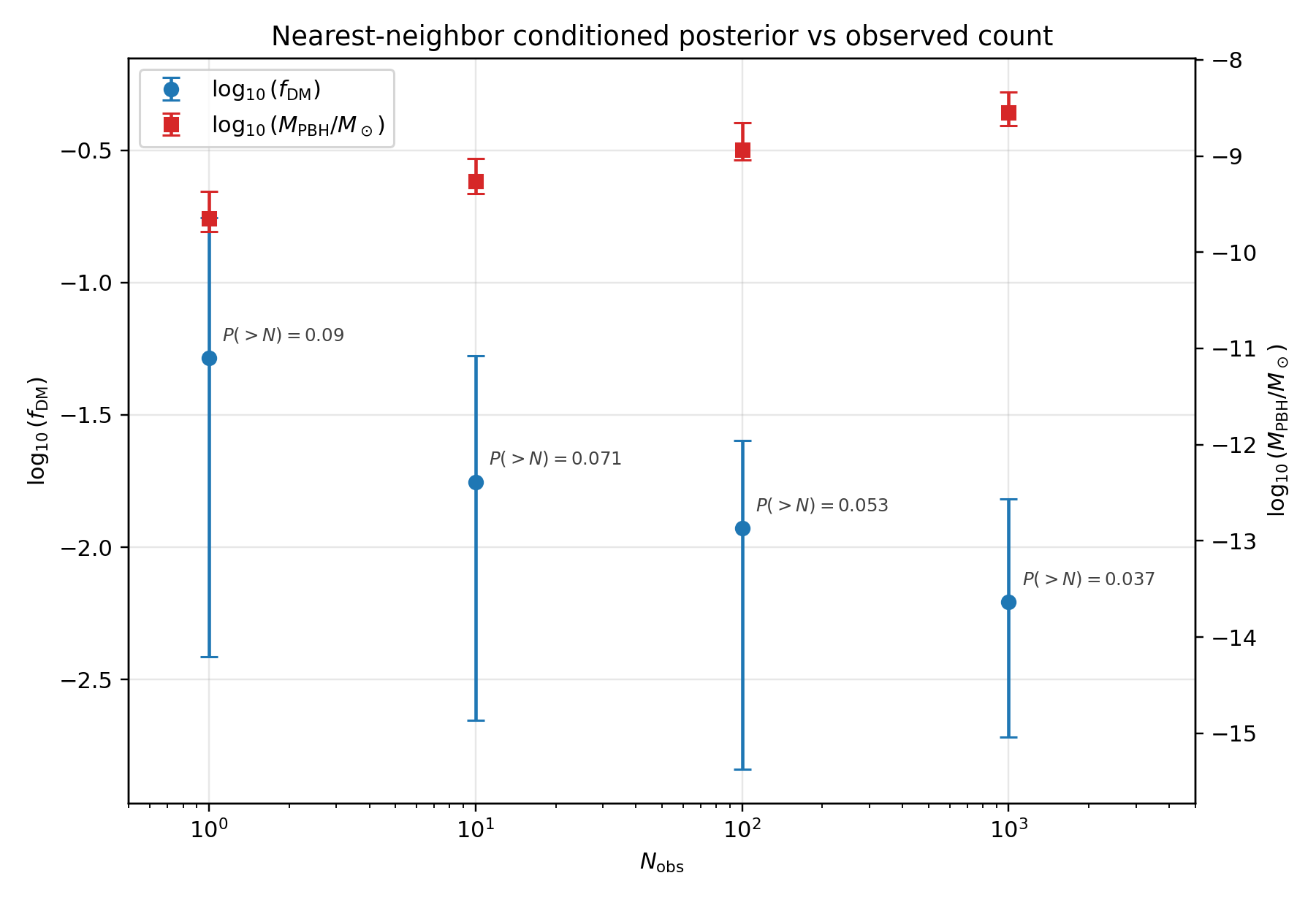}

\vspace{0.6em}

\includegraphics[width=1\columnwidth]{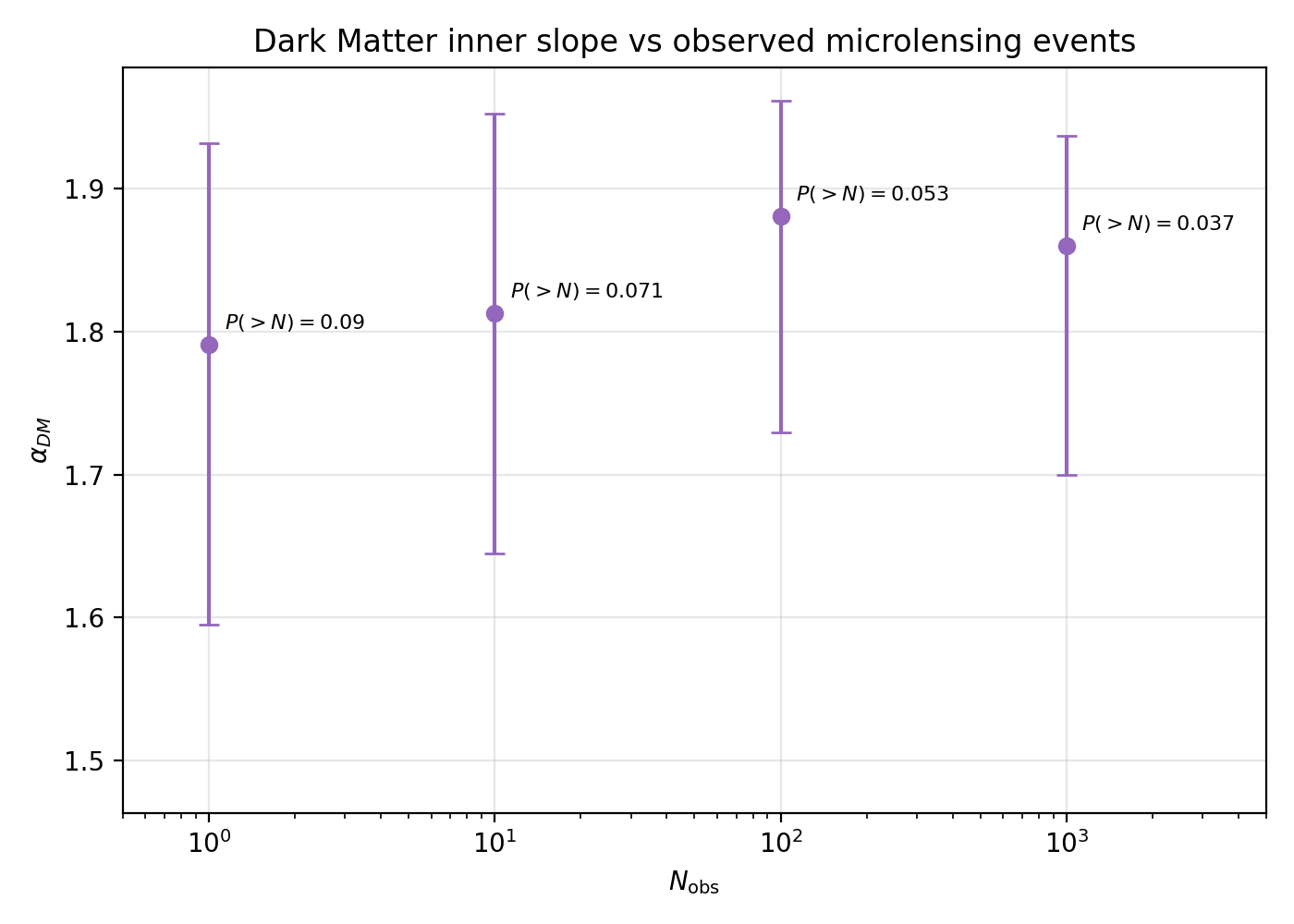}
\caption{Posterior constraints conditioned on an assumed Roman microlensing count $N_{\rm obs}$. For each $N_{\rm obs}$ we reweight posterior samples by the Poisson likelihood ${\rm Pois}(N_{\rm obs}\,|\,N_{\rm det})$ and report the weighted 16th and 84th percentiles (error bars) of $\log_{10}(M_{\rm PBH}/M_\odot)$ and $\log_{10}(f_{\rm DM})$ (top), and the inner dark-matter slope $\alpha$ (bottom). Conditioning primarily constrains $M_{\rm PBH}$ through the steep mass dependence of the event rate, while $f_{\rm DM}$ remains more weakly constrained due to its role as an overall normalization and its degeneracy with $M_{\rm PBH}$; $\alpha$ is comparatively insensitive in this regime.}
\label{fig:PBH_lensing_Roman_conditioned}
\end{figure}

This procedure yields Monte Carlo samples from the Roman-conditioned posterior and quantifies how future microlensing measurements would refine the allowed PBH parameter space. As shown in Fig.~\ref{fig:PBH_lensing_Roman_conditioned}, the predicted Roman event yield exhibits a steep dependence on $M_{\rm PBH}$ across the mass range supported by the G-object posterior. The function $N_{\rm det}(M_{\rm PBH})$ varies much more rapidly with mass than with any other parameter, so conditioning on a given observed event count $N_{\rm obs}$ tightly constrains $M_{\rm PBH}$ within a comparatively narrow interval.

By contrast, $f_{\rm DM}$ enters primarily as an overall linear normalization of the PBH lens density,
\begin{equation*}
\lambda(\boldsymbol{\theta}) \propto f_{\rm DM},
\end{equation*}
at fixed mass. Variations in $f_{\rm DM}$ therefore rescale the expected yield without significantly altering its shape as a function of $M_{\rm PBH}$, leading to a broader allowed range compared to the sharply localized mass constraints.

The Roman yield curve is non-monotonic in $M_{\rm PBH}$: the event rate increases with mass up to a characteristic scale and subsequently declines. In principle, this behavior allows two distinct mass solutions to reproduce a given $N_{\rm obs}$ at fixed $f_{\rm DM}$. However, the higher-mass branch lies outside the dominant support of the  posterior and is therefore effectively excluded in the joint inference. As a result, the Roman-conditioned posterior selects a single, narrow mass region.

Finally, over the broad mass interval permitted by the GC constraints, the predicted yield shows only weak sensitivity to the inner slope $\alpha_{\rm DM}$. The uncertainties shown in Fig.~\ref{fig:PBH_lensing_Roman_conditioned} therefore largely reflect the intrinsic width of the G-object posterior. In practice, the Roman likelihood primarily sharpens the inference in the $(M_{\rm PBH}, f_{\rm DM})$ plane, while $\alpha_{\rm DM}$ remains dominated by its non-microlensing posterior support.

\subsection{Tidal disruption signatures of G objects in standard and PBH--halo scenarios}
\label{subsec:G_objects_tidal}

In both standard and PBH capture scenarios, G objects are expected to survive pericentre as a compact source, consistent with the observed persistence of G2 in L$'$ and its continued Keplerian motion over more than a decade 
\cite{Witzel:2014G2,Plewa:2017G2}. However, detailed predictions for morphology, line emission, continuum variability, and orbital perturbations differ in important ways, which we summarize below.

\subsubsection{Qualitative expectations near pericentre}

In the \emph{standard} (dust--enshrouded star / merger) case, the extended envelope can be tidally sheared into an elongated structure in position--velocity space as the object crosses pericentre. The recombination--line emission (e.g.\ Br$\gamma$ at $2.166\,\mu{\rm m}$) traces this sheared gas, producing a velocity gradient and extended emission along the orbit; observations of G2 prior to and during pericentre indeed show such tidal stretching in Br$\gamma$ while the source remains unresolved in L$'$.%
\cite{Gillessen:2013G2,Witzel:2014G2,Plewa:2017G2} The envelope may also suffer partial ram--pressure stripping by the ambient medium. A simple estimate of the stripped mass over an interaction time $t_{\rm int}$ is
\begin{equation}
  \Delta M \sim \pi R_{\rm env}^2 \,\rho_{\rm amb}(r_{\rm p})\, v_{\rm rel}\, t_{\rm int}\,,
\end{equation}
where $R_{\rm env}$ is the envelope radius, $\rho_{\rm amb}(r_{\rm p})$ is the ambient density at pericentre distance $r_{\rm p}$, and $v_{\rm rel}$ is the relative velocity. For $R_{\rm env}\sim {\rm few~AU}$ and densities consistent with constraints from S2 and G2,%
\cite{Hosseini:2020S2density,Gillessen:2018Drag} this stripping is partial rather than catastrophic.

In the \emph{PBH + gas halo} case, the bound halo is more compact and more strongly gravitationally bound to the central mass $M_\bullet$; only the outermost layers can be stripped. If we approximate the halo as a sphere of radius $R_{\rm h}$ and mass $M_{\rm h}\ll M_\bullet$, the condition for substantial ram--pressure stripping at radius $r_{\rm p}$ is roughly
\begin{equation}
  \rho_{\rm amb}(r_{\rm p})\,v_{\rm rel}^2 \gtrsim \frac{G M_\bullet M_{\rm h}}{R_{\rm h}^4}\,,
\end{equation}
so that for AU--scale halos and $M_\bullet\gtrsim 10^{-2}$--$1\,M_\odot$ only the outer layers are affected for plausible Galactic Centre densities. The corresponding drag acceleration is
\begin{equation}
  a_{\rm drag} \sim \frac{C_{\rm D}\,\rho_{\rm amb}\,v^2\,\pi R_{\rm h}^2}{M_\bullet+M_{\rm h}}\,,
\end{equation}
with $C_{\rm D}\sim\mathcal{O}(1)$. For $R_{\rm h}\sim{\rm AU}$ and the densities inferred from G2 and S2, this acceleration is too small to produce detectable deviations from a purely Keplerian orbit over the current observational baseline, in agreement with the non--detection of strong drag in G2 and with the persistence of its L$'$ brightness.%
\cite{Plewa:2017G2,Hosseini:2020S2density,Gillessen:2018Drag}

\subsubsection{Observational discriminants}

% Put this near your table (or in the preamble)
\newcommand{\wrapcell}[2]{%
  \parbox[t]{#1}{\raggedright\strut #2\strut\vspace{6pt}}}

\begin{table*}[t]
  \caption{Qualitative diagnostics for G objects in the standard stellar--envelope vs PBH + gas--halo scenarios near pericentre.}
  \label{tab:Gobject_diagnostics}
  \centering
  \renewcommand{\arraystretch}{3.0}

  \begin{ruledtabular}
  \begin{tabular*}{\textwidth}{@{\extracolsep{\fill}}ccc}
    \hline
    \wrapcell{0.22\textwidth}{Observable} &
    \wrapcell{0.37\textwidth}{Standard G object (dust--enshrouded star / merger)} &
    \wrapcell{0.37\textwidth}{PBH + gas halo (sub--solar--mass PBH + atmosphere)} \\
    \hline
    \wrapcell{0.22\textwidth}{Morphology in Br$\gamma$} &
    \wrapcell{0.37\textwidth}{Strong tidal stretching along the orbit; extended position--velocity structure; possible leading/trailing tail.} &
    \wrapcell{0.37\textwidth}{Compact core plus thin stripped sheath or bow--shock--like feature; much weaker global shear of the emitting region.} \\
    \hline
    \wrapcell{0.22\textwidth}{Br$\gamma$ line evolution} &
    \wrapcell{0.37\textwidth}{Line width and PV extent increase near pericentre; luminosity roughly constant or mildly varying over years.} &
    \wrapcell{0.37\textwidth}{Spatially compact emission; modest brightening possible near pericentre, but PV extent remains limited.} \\
    \hline
    \wrapcell{0.22\textwidth}{L$'$ continuum evolution} &
    \wrapcell{0.37\textwidth}{Warm--dust L$'$ luminosity approximately constant over a decade, with at most modest changes around pericentre, consistent with a dust shell around a stellar source.} &
    \wrapcell{0.37\textwidth}{Also roughly constant L$'$ luminosity; dust in the bound halo reprocesses accretion and external radiation; strong flares from Sgr~A* are not required and are observationally absent.} \\
    \hline
    \wrapcell{0.22\textwidth}{Orbital dynamics} &
    \wrapcell{0.37\textwidth}{Keplerian motion within uncertainties; weak or undetectable drag; multiple G objects share similar phenomenology.} &
    \wrapcell{0.37\textwidth}{Keplerian motion to even higher precision; drag further suppressed for compact halos; deviations from a test--particle orbit are expected to be below current detection limits.} \\
    \hline
    \wrapcell{0.22\textwidth}{Post--pericentre evolution} &
    \wrapcell{0.37\textwidth}{Envelope re--compacts on multi--year timescales; source remains detectable and stellar--like in dynamics.} &
    \wrapcell{0.37\textwidth}{Atmosphere recircularizes around PBH; any stripped material forms a diffuse tail; compact core remains as a long--lived IR source.} \\
    \hline
  \end{tabular*}
  \end{ruledtabular}
\end{table*}

Table~\ref{tab:Gobject_diagnostics} summarizes the main expected differences between the two classes of models.

\subsubsection{PBH + halo parametrization and scaling relations}

For quantitative modelling of a PBH--driven G object, we adopt the following parametrization:
\begin{equation*}
\begin{gathered}
M_\bullet = \text{mass of the compact object (PBH--seeded BH)},\\
M_{\rm h} = \text{mass of the bound gaseous halo},\\
R_{\rm h} = \text{outer radius of the halo},\\
\rho(r) = \rho_0 \left(\frac{r}{R_{\rm h}}\right)^{-\alpha}\,, \quad 0 \le r \le R_{\rm h},
\end{gathered}
\end{equation*}

with $\alpha$ a density--slope parameter and $\rho_0$ fixed by $M_{\rm h}$,
\begin{equation}
  M_{\rm h} = 4\pi \int_0^{R_{\rm h}} r^2 \rho(r)\,{\rm d}r
  = \frac{4\pi \rho_0 R_{\rm h}^3}{3-\alpha}\,,
  \quad (\alpha<3)\,.
\end{equation}
Assuming a fully ionized gas with mean molecular weight $\mu$, the electron density profile is $n_e(r) \simeq \rho(r)/( \mu m_p )$.

\begin{figure}[t]
\centering
\includegraphics[width=1\linewidth]{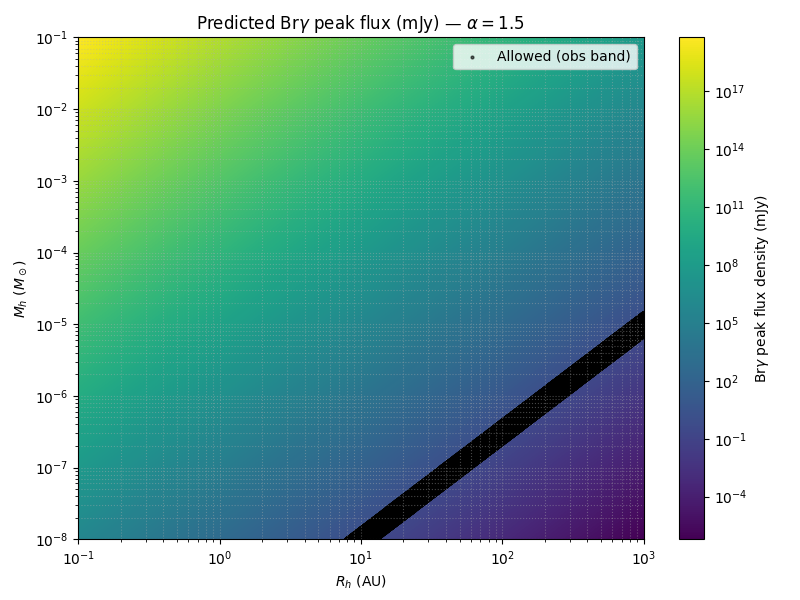}
\caption{Color-map showing how the mass of the halo varies with the radius of the dust envelope in the density bounded regime. The black band represents the current Br$\gamma$ observations which can be used to constrain the compactness of the G objects
}
\label{fig:Brycolormap}
\end{figure}

For case--B recombination in a region with roughly uniform ionization fraction $x_e$ and electron temperature $T_e$, the Br$\gamma$ luminosity scales as
\begin{equation}
  L_{{\rm Br}\gamma} \simeq h\nu_{{\rm Br}\gamma}\,\alpha_{\rm eff}({\rm Br}\gamma,T_e)
  \int n_e^2(r)\,{\rm d}V
  \propto \frac{M_{\rm h}^2}{R_{\rm h}^3}\, f(\alpha)\,,
\end{equation}
where $\alpha_{\rm eff}({\rm Br}\gamma,T_e)$ is the effective recombination coefficient and $f(\alpha)$ is an order--unity function encapsulating the density profile:
\begin{equation}
  f(\alpha) \equiv 
  \frac{\int_0^{R_{\rm h}} r^2 \left(r/R_{\rm h}\right)^{-2\alpha}\,{\rm d}r}
       {\left[\int_0^{R_{\rm h}} r^2 \left(r/R_{\rm h}\right)^{-\alpha}\,{\rm d}r\right]^2}
  \sim \mathcal{O}(1)\,.
\end{equation}
Thus, neglecting order--unity factors,
\begin{equation}
  L_{{\rm Br}\gamma} \propto M_{\rm h}^2\,R_{\rm h}^{-3}\,,
\end{equation}
so that for fixed $L_{{\rm Br}\gamma}$ the allowed $(M_{\rm h},R_{\rm h})$ combinations lie approximately along $M_{\rm h}\propto R_{\rm h}^{3/2}$. In Fig \ref{fig:Brycolormap} we plot the allowed range for $M_{h}, \,R_h$.

\begin{equation}
  L_{\rm heat} \sim 16\pi R_d^2\,\sigma_{\rm SB}\,T_d^4\,,
\end{equation}
and hence
\begin{equation}
  L_{L'} \propto C_f \left(1-e^{-\tau_{L'}}\right) L_{\rm heat}\,.
\end{equation}
In an accretion--powered scenario, $L_{\rm heat} \sim \eta \dot{M}_\bullet c^2$, with $\eta$ an efficiency and $\dot{M}_\bullet$ set either by Bondi--like inflow from the halo or by externally supplied gas. For Bondi scaling inside the halo,
\begin{equation}
  \dot{M}_\bullet \sim 4\pi \lambda \frac{(G M_\bullet)^2 \rho(R_{\rm B})}{c_s^3}
  \propto M_\bullet^2\,\rho_0\,c_s^{-3}\,,
\end{equation}
so that, in the simplest case,
\begin{equation}
  L_{L'} \propto M_\bullet^2\,\rho_0 \propto M_\bullet^2\,\frac{M_{\rm h}}{R_{\rm h}^3}\,,
\end{equation}
up to order--unity factors and the unknown efficiency $\eta$.

These proportionalities provide a convenient fitting framework: given observed $L_{{\rm Br}\gamma}$ and $L_{L'}$ (and their temporal evolution), one can constrain combinations of $(M_\bullet,M_{\rm h},R_{\rm h},\alpha)$ and the heating efficiency. In particular,
\begin{align}
  L_{{\rm Br}\gamma} &\propto \left(\frac{M_{\rm h}}{R_{\rm h}^{3/2}}\right)^2, \\
  L_{L'} &\propto M_\bullet^2\,\frac{M_{\rm h}}{R_{\rm h}^3}\,,
\end{align}
so that, to leading order, the ratio
\begin{equation}
  \frac{L_{L'}}{L_{{\rm Br}\gamma}^{1/2}} \propto M_\bullet^2\,M_{\rm h}^{-1/2}\,R_{\rm h}^{-3/2}
\end{equation}
is sensitive to the compactness $M_\bullet/R_{\rm h}$ of the central object and the relative mass in the halo.

%\subsubsection{Figure placeholder}

\begin{figure*}
  \centering
  \includegraphics[width=0.7\textwidth]{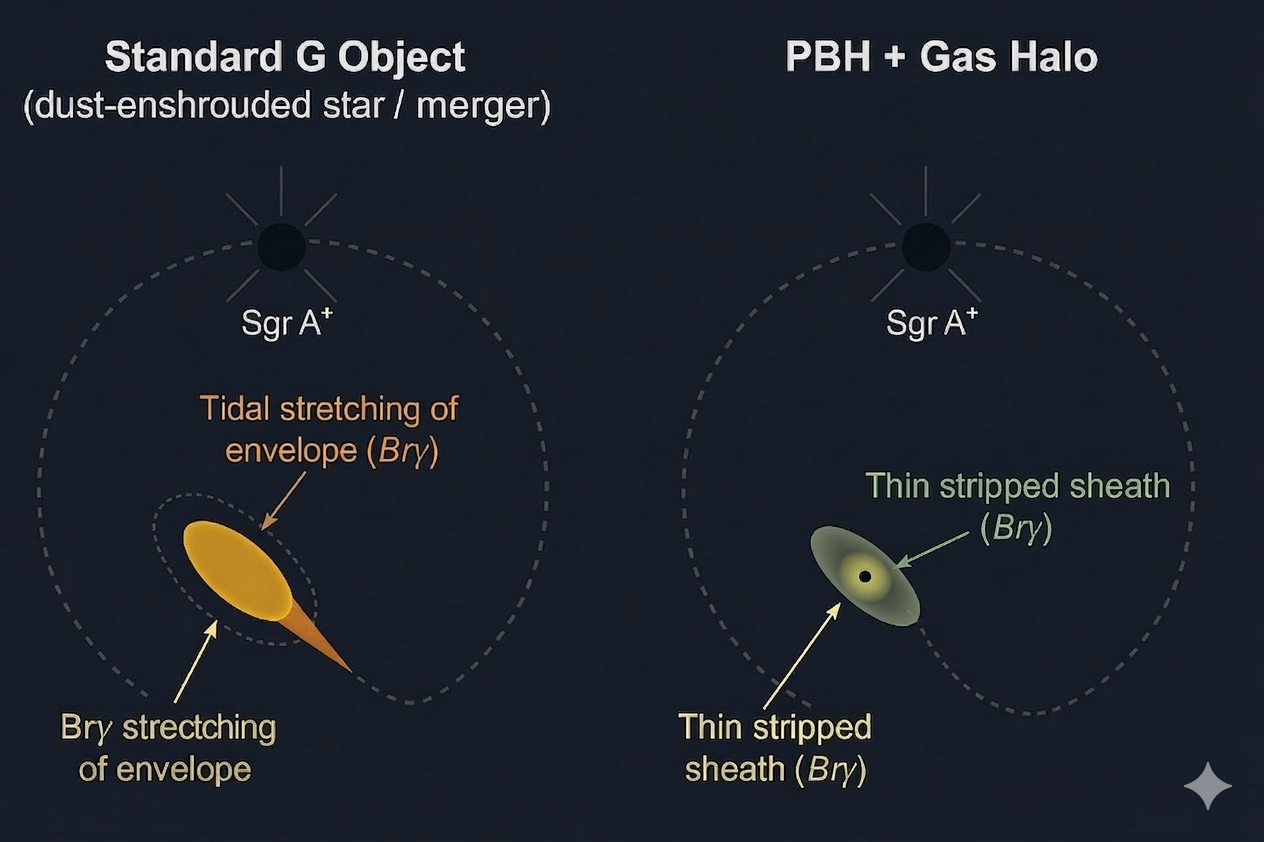}
  \caption{Schematic comparison of the tidal response of a standard G object (left; dust--enshrouded star or merger product) and a PBH + gas halo (right) during pericentre passage around Sgr~A*. In the stellar--envelope case, the extended circumstellar material is significantly sheared, producing an elongated Br$\gamma$ structure with a possible leading/trailing tail, while the L$'$ continuum traces a more compact dust core. In the PBH--halo case, the gravitationally bound atmosphere remains compact, with only its outer layers forming a thin stripped sheath; both the Br$\gamma$ and L$'$ emission stay closely associated with a point--like core, and hydrodynamic drag on the orbit is strongly suppressed.}
  \label{fig:Gobject_schematic}
\end{figure*}

In Fig.~\ref{fig:Gobject_schematic} we schematically compare the expected morphology and emission regions in the two scenarios. In the standard case, the extended envelope is visibly sheared along the orbit, with Br$\gamma$ tracing a tidal tail; in the PBH--halo case, the Br$\gamma$ emission is concentrated in a compact core with a thin, low--surface--brightness sheath, while the L$'$ continuum remains tightly bound to the central object in both scenarios.

In practice, the most powerful discriminants are: (i) high--resolution position--velocity mapping of Br$\gamma$ around pericentre, (ii) precise measurements of secular changes in orbital elements (probing drag and thus effective cross--section), and (iii) long--term monitoring of the L$'$ spectral energy distribution and variability. Existing data already exclude pure, unbound gas clouds and support a \emph{bound core} in both scenarios.%
\cite{Gillessen:2013G2,Witzel:2014G2,Ciurlo:2020Nature,Plewa:2017G2,Flores:2023PRD}
\section{Discussion: Observational Tests of the PBH--NS Remnant Scenario}
\label{sec:discussion}

A primary goal of this work is to identify observational tests that can distinguish the hypothesis that G objects are primordial-black-hole--induced neutron-star remnants from more conventional stellar interpretations. In this section, we summarize the key diagnostics developed throughout the paper and emphasize how they jointly provide a coherent and falsifiable framework.

\subsection{Infrared continuum and recombination-line signatures}

The defining observational properties of G objects are established in Sec.~\ref{sec:observational_status}, including their compact Br$\gamma$ emission, extreme $K'-L'$ colors, and survival through repeated periapse passages. We showed that these features arise naturally if the central object is a low-mass black hole surrounded by a compact envelope of gas and dust that reprocesses accretion luminosity.

A particularly important observational constraint arises from the infrared emission of G objects. The elevated dust temperatures inferred for these sources cannot be explained by external irradiation alone, indicating the presence of an internal heating mechanism. In this framework, accretion onto a compact object embedded within a dusty envelope provides a natural explanation for the observed infrared properties, with luminosities consistent with radiatively inefficient accretion flows. This behavior is difficult to reproduce in stellar-envelope models, where the infrared emission is tied primarily to stellar luminosity and disk geometry.

Similarly, the observed Br$\gamma$ luminosities can be reproduced by photoionization from radiatively inefficient accretion flows in both the ionization-bounded and density-bounded regimes. In this picture, the line emission traces the ionizing photon budget rather than stellar ultraviolet radiation, leading to characteristic correlations between recombination-line strength, accretion rate, and envelope mass.

\subsection{Radio continuum and recombination lines}

Ionized gas bound to PBH--NS remnants produces thermal free--free emission with nearly flat radio spectra and weak hydrogen radio recombination lines. For plausible envelope densities and sizes, the expected flux densities fall in the sub-mJy regime and are accessible to current and next-generation radio facilities.

These radio properties provide a powerful discriminant. Young stellar objects and stellar winds typically exhibit rising radio spectra, while synchrotron sources display steep non-thermal slopes. Detection of flat-spectrum radio emission spatially coincident with infrared-identified G objects would therefore strongly favor an ionized-envelope interpretation. Hydrogen recombination lines such as H30$\alpha$ or H41$\alpha$ would further probe the gas kinematics and physical conditions, enabling direct comparisons with orbital parameters inferred from infrared spectroscopy.

\subsection{X-ray faintness and accretion inefficiency}

An important indirect test of the PBH--NS remnant scenario is the expected X-ray faintness of G objects. As we argued before, the accretion rates consistent with the infrared and recombination-line data lie deeply in the radiatively inefficient regime. This naturally explains the absence of persistent or transient X-ray counterparts despite the presence of compact objects.

This behavior contrasts with expectations for accreting stellar-mass black holes in binary systems and supports an interpretation in which G objects are isolated remnants embedded in low-density envelopes rather than conventional X-ray binaries.

\subsection{Population-level tests and the missing pulsar problem}

Beyond individual objects, the PBH--NS conversion hypothesis makes distinctive population-level predictions. While this mechanism can produce a population of G objects through the destruction of isolated neutron stars, our analysis shows that it is not efficient enough to account for the observed deficit of ordinary radio pulsars in the Galactic Center. Instead, the missing-pulsar constraint primarily acts to limit the underlying population of isolated neutron stars available for conversion.

In this framework, the absence of detected canonical pulsars does not require significant PBH-induced depletion, but rather informs the normalization and demographics of the neutron-star population, particularly the fraction that is isolated and radio-active. Consequently, the production rate of G objects is tied not only to PBH physics but also to the availability of isolated neutron stars, which may be reduced by evolutionary history or binary populations.

Future improvements in pulsar surveys toward the Galactic Center therefore remain a crucial indirect probe. A continued lack of canonical pulsar detections would further constrain the population of isolated neutron stars, while an expanding census of G objects would help determine whether PBH-induced conversion operates within this limited reservoir. Together, these observations can test the internal consistency of the PBH--NS remnant scenario without requiring it to resolve the missing pulsar problem itself.

\subsection{Microlensing probes of the underlying PBH population}

Gravitational microlensing offers a complementary and independent probe of the PBH population invoked in this scenario. Upcoming time-domain surveys such as the Roman Space Telescope Galactic Bulge Time-Domain Survey will have sensitivity to PBH masses overlapping the preferred region of parameter space inferred from the joint analysis.

A detection or strong upper limit on short-timescale microlensing events would therefore provide an external consistency check on the PBH interpretation of G objects. Importantly, this test relies only on gravitational effects and does not depend on modeling the complex baryonic environment of the Galactic Center.

\subsection{Outlook}

Taken together, the observational tests summarized above demonstrate that the PBH--NS remnant scenario is both predictive and falsifiable. Infrared spectroscopy, sensitive radio observations, pulsar surveys, and microlensing experiments probe distinct aspects of the model, and their combination offers a robust path toward confirming or excluding this interpretation. Regardless of the outcome, G objects emerge as a valuable laboratory for studying compact-object physics, extreme Galactic environments, and the small-scale properties of dark matter.

\section{Conclusions}
\label{sec:conclusions}

The nature of the G objects in the Galactic Center continues to pose a challenge to standard astrophysical interpretations. Their compactness, persistent Br$\gamma$ emission, extreme infrared colors, and survival through repeated close periapse passages around Sgr~A$^\ast$ are difficult to reconcile with models based solely on unbound gas clouds, ordinary stars, or transient stellar merger remnants. Motivated by these tensions, we have explored the hypothesis that G objects are the long-lived remnants of neutron stars converted into low-mass black holes through the capture of primordial black holes (PBHs).

In this work, we developed a population-level framework that connects PBH--neutron-star conversion physics to the observable demographics of the Galactic Center. By modeling the neutron-star distribution, the inner dark-matter density profile, and the PBH mass and abundance, we showed that the expected number and spatial distribution of converted remnants can naturally reproduce the observed G-object population. Within the same framework, we found that the NS-PBH conversion scenario does not occur efficiently enough as to deplete an underlying CP population and thus cannot be a solution to the long standing galactic center missing pulsar problem.

We noted that millisecond pulsars (MSPs) are not consistent with the observed properties of the G-objects. However,  MSPs could be significantly overrepresented in the Galactic Center, potentially comprising a large fraction of the neutron star population. In such a scenario, if primordial black holes (PBHs) are capable of destroying neutron stars, MSPs warrant particular consideration. Owing to their long lifetimes ($\sim 10^9\,\mathrm{yr}$), MSPs could experience substantial cumulative capture probabilities. If their PBH capture rates are comparable to those of canonical pulsars, a significant fraction of the MSP population could be converted over their lifetimes, offering an additional perspective on the missing pulsar problem.

\begin{figure}[t]
\centering
\includegraphics[width=1\columnwidth]{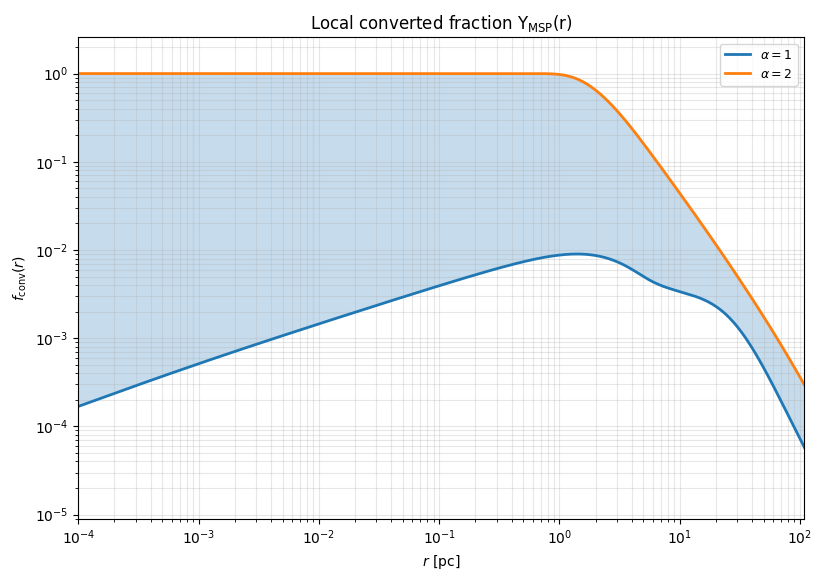}
\caption{Converted fraction of millisecond pulsars (MSPs) due to PBH capture as a function of Galactocentric radius. Owing to their long lifetimes ($\sim 10^9\,\mathrm{yr}$), MSPs can reach high conversion fractions in the innermost regions, potentially approaching unity within $\sim 1\,\mathrm{pc}$. At larger radii ($\gtrsim 10\,\mathrm{pc}$), where surveys are highly sensitive, the converted fraction declines, providing observational constraints.}
\label{fig:MSP_conv}
\end{figure}

In Fig.~\ref{fig:MSP_conv} we illustrate the converted fraction for MSPs. In the most extreme cases, the MSP population could be nearly completely converted within $\sim 1\,\mathrm{pc}$ of the Galactic Center. However, current pulsar surveys have achieved high sensitivities out to radii of $\gtrsim 10\,\mathrm{pc}$, placing important constraints on such scenarios. A detailed treatment of MSP capture rates, spatial distributions, and survey selection effects is therefore required to assess this possibility more rigorously and is left to future work.

If the neutron star binary fraction in the Galactic Center is close to unity, a large fraction of neutron stars may be recycled into MSPs, leaving comparatively few isolated neutron stars. Since PBH capture and subsequent conversion into G-objects requires isolated neutron stars, such a scenario could limit the available reservoir for PBH-induced conversions. In this limit, there may be insufficient isolated neutron stars to account for the observed G-object population through PBH capture alone.

Additionally, if MSPs are efficiently converted within their lifetimes, the resulting events may have broader astrophysical implications. As shown by \citet{Fuller_2017}, neutron star--PBH conversion events could produce r-process nucleosynthesis in the ejecta. If $\mathcal{O}(10^5)$ such conversions occurred over Galactic history, they could potentially account for a substantial fraction of the Milky Way's r-process material. Quantifying this connection within a self-consistent Galactic model/Missing Pulsar problem remains an important direction for future work.

A key result of our analysis is that the G-object population is expected to be highly incomplete, with current observations plausibly probing only a few percent of the underlying parent population. When this incompleteness is taken into account, the PBH-induced conversion scenario is internally consistent across a wide region of parameter space that remains compatible with existing astrophysical and cosmological constraints. The preferred parameter regions favor steep inner profiles for both the neutron-star and dark-matter distributions, reflecting the strong environmental dependence of compact-object capture processes in the Galactic Center.

Beyond population statistics, we identified a suite of multi-wavelength observational diagnostics that can decisively test this scenario. These include characteristic infrared colors and recombination-line luminosities from gas bound to low-mass black holes, flat-spectrum thermal radio emission from ionized envelopes, the absence of stellar photospheric features, and a potentially detectable microlensing signal from the underlying PBH population in upcoming time-domain surveys. Importantly, these signatures differ qualitatively from those expected in conventional stellar-envelope or merger-based models.

We emphasize that our goal is not to claim that G objects must originate from PBH--neutron-star interactions, but rather to demonstrate that this hypothesis provides a coherent and predictive framework that unifies several otherwise disconnected Galactic Center puzzles. If confirmed, G objects would represent a novel class of compact remnants and a new astrophysical probe of dark matter on small scales. Future high-resolution infrared spectroscopy, sensitive radio observations, and microlensing measurements—particularly with next-generation facilities—will be crucial in determining whether this scenario plays a role in shaping the compact-object population of the Galactic Center.

\section*{Acknowledgments}
This work is partly supported by the U.S. Department of Energy
grant number de-sc0010107 (SP).

D.M.Z was supported by the U.S. National Science Foundation Grant PHY-1913693

\bibliography{newbib_clean}% Produces the bibliography via BibTeX.

\end{document}